\documentclass[a4paper,11pt]{article}

\usepackage{amsmath}
\usepackage{graphicx}
\usepackage{amsfonts}
\usepackage{amssymb}

\newtheorem{theorem}{Theorem}[section]
\newtheorem{rema}{Remark}[section]
\newtheorem{assump}{Assumption}[section]
\newtheorem{propo}[rema]{Proposition}
\newtheorem{defi}[rema]{Definition}
\newtheorem{lemma}[rema]{Lemma}
\newtheorem{corol}[rema]{Corollary}

\newcommand{\sect}[1]{\setcounter{equation}{0}\section{#1}}

\newcommand{\bc}{\begin{center}}
\newcommand{\ec}{\end{center}}
\def\ba#1{\begin{array}{#1}\displaystyle}
\newcommand{\ea}{\end{array}}
\newcommand{\z}{\\[2mm] \displaystyle}

\newcommand{\beq}{\begin{equation}}
\newcommand{\eeq}{\end{equation}}
\newcommand{\beqa}{\begin{eqnarray}}
\newcommand{\eeqa}{\end{eqnarray}}
\newcommand{\no}{\nonumber}
\newcommand{\n}{\nonumber\\}
\newcommand{\bi}{\begin{itemize}}
\newcommand{\ei}{\end{itemize}}
\def\lt#1{\left#1}
\def\rt#1{\right#1}
\def\t#1{\tilde{#1}}

\def\b#1{\bar{#1}}
\def\frc#1#2{\frac{#1}{#2}}

\newcommand{\p}{\partial}

\newcommand{\bra}{\langle}
\newcommand{\ket}{\rangle}
\newcommand{\Z}{{\mathbb{Z}}}

\newcommand{\R}{{\mathbb{R}}}
\newcommand{\C}{{\mathbb{C}}}
\newcommand{\hC}{{\hat{\mathbb{C}}}}
\newcommand{\uH}{{\mathbb{H}}}

\newcommand{\uD}{{\mathbb{D}}}

\newcommand{\Or}{{\cal O}}

\newcommand{\ep}{\epsilon}
\newcommand{\varep}{\varepsilon}

\newcommand{\id}{{\rm id}}

\newcommand{\halmos}{\rule{1ex}{1.4ex}}
\newcommand{\eproof}{\hspace*{\fill}\mbox{$\halmos$}}
\newcommand{\proof}{{\em Proof.\ }}

\newcommand{\gl}{{\tt G}}
\newcommand{\kl}{{\tt K}}
\newcommand{\dom}{{\Upsilon}}
\newcommand{\se}{\Omega}
\newcommand{\ev}{{\cal E}}

\newcommand{\tou}{{\cal X}}

\newcommand{\sym}{{\tt S}}
\newcommand{\Preg}{P^{ren}}

\newcommand{\supp}{{\rm supp}}

\def\cl#1{\overline{#1}}
\newcommand{\chrg}{\Gamma}
\newcommand{\dd}{{\rm d}}
\newcommand{\bd}{\b{{\rm d}}}

\begin{document}

\begin{titlepage}

\begin{center}
{\Large {\bf
Conformal loop ensembles and the stress-energy tensor.

II. Construction of the stress-energy tensor}

\vspace{1cm}

Benjamin Doyon}

Department of Mathematical Sciences, Durham University\\
South Road, Durham DH1 3LE, UK\\
email: benjamin.doyon@durham.ac.uk

\end{center}

\vspace{1cm}

\noindent This is the second part of a work aimed at constructing the stress-energy tensor of conformal field theory (CFT) as a local ``object'' in conformal loop ensembles (CLE). This work lies in the wider context of re-constructing quantum field theory from mathematically well-defined ensembles of random objects. In the present paper, based on results of the first part, we identify the stress-energy tensor in the dilute regime of CLE. This is done by deriving both its conformal Ward identities for single insertion in CLE probability functions, and its properties under conformal transformations involving the Schwarzian derivative. We also give the one-point function of the stress-energy tensor in terms of a notion of partition function, and we show that this agrees with standard CFT arguments. The construction is in the same spirit as that found in the context of SLE$_{8/3}$ by the author, Riva and Cardy (2006), which had to do with the case of zero central charge. The present construction generalises this to all central charges between 0 and 1, including all minimal models. This generalisation is non-trivial: the application of these ideas to the CLE context requires the introduction of a {\em renormalised probability}, and the derivation of the transformation properties and of the one-point function do not have counterparts in the SLE context.

\vfill

{\ }\hfill 19 October 2009

\end{titlepage}

\tableofcontents

\sect{Introduction}

Quantum field theory (QFT) is one of the most successful theory of modern physics. It describes the full universal, large-distance behaviour of statistical systems near thermal critical points, and of quantum systems near quantum critical points (the scaling limit). It also provides a powerful description of relativistic quantum particles.

Two-dimensional conformal field theory (CFT), describing the critical point itself and displaying scale invariance, constitutes a particular family of QFT models which enjoy somewhat more accurate mathematical descriptions. The corner stone of many of these descriptions is the stress-energy tensor (also called the energy-momentum tensor). Besides its mathematical properties, this object is physically the most important, and has clear interpretations in all ways of understanding QFT. From the viewpoint of statistical models, this is a local fluctuating tensor variable that describes changes in the (Euclidean-signature) metric. From the viewpoint of quantum chains, it is perhaps more naturally seen as grouping together the conserved currents underlying space translation invariance (stress) and time translation invariance (energy). In a similar spirit, from the viewpoint of relativistic particles, it is a local measure of the flow of momentum and energy.

The study of the stress-energy tensor in CFT gives rise to the full algebraic construction of CFT (see the lecture notes \cite{Gins}, or the standard textbook \cite{DFMS97} and references therein). In general, a QFT model can be defined algebraically by providing a Hilbert space (in a given quantisation direction) as a module for the space-time symmetry algebra, along with the action of the stress-energy tensor. The full construction of a local sector of the QFT model is then obtained by constructing all mutually local field-operators that are also local with respect to the stress-energy tensor. In CFT, the space-time symmetry algebra is usually taken as the algebra of the generators of the quantum-mechanically broken local conformal symmetry: two independent copies of the Virasoro algebra -- although only a small subalgebra describes actual {\em symmetries}. This is useful, because the Hilbert space can be taken as a module for these two independent copies of the Virasoro algebra, and the stress-energy tensor is expressed linearly in terms of Virasoro elements. The central charge of the Virasoro algebra and a choice of two-copy Virasoro module then fully defines the model. The complete mathematical framework where these ideas are realised is that of vertex operator algebras (see, for instance, \cite{LL04}).

Besides the powerful algebraic description of QFT, one often refers, although usually more informally, to probabilistic descriptions: fluctuating fields, particle trajectories, etc. It is fair to say that these descriptions are not as well developed mathematically, although they provide a more global view on QFT, facilitating the treatment of topological effects and without the need for an explicit quantisation direction. Recently, Sheffield and Werner developed a new, consistent probabilistic description of CFT: that of conformal loop ensembles (CLE) \cite{W05a,Sh06,ShW07}. Loosely speaking, these constitute measures on ensembles of non-crossing loops, where the loops could be thought of as iso-height lines of fluctuating fields. These loop descriptions have the advantage of being much nearer to statistical models underlying CFT: fluctuating loops are, in a sense, the objects with a proper scaling limit (see the discussion in \cite{I}). This is a giant step towards a better understanding of CFT and QFT more generally, from many viewpoints: having a mathematically consistent probabilistic theory of QFT, connecting QFT to underlying discrete models, and getting a full description of the true scaling objects.

The present paper is the second part of a work started in \cite{I}. The goal of this work is to construct the stress-energy tensor in CLE, and derive its main properties at the basis of the algebraic description of CFT. Since the stress-energy tensor has clear interpretations in the three physical paths to QFT described above, its identification in CLE provides a better physical understanding of the fluctuating CLE loops. Moreover, the algebraic description of CFT is until now by far the most useful for making non-trivial predictions, whereas only CLE can be mathematically shown to occur in the scaling limit of many statistical models \cite{Smi}. Connecting algebraic CFT to CLE could provide a mathematical path from statistical models to the powerful algebraic machinery, something which has never been done for any non-trivial QFT.

In \cite{I}, we provided an introduction and overview of CLE and its connection to CFT, and we developed new notions in the CLE context, obtaining some basic results about them. In the present paper, we use these notions and basic results in order to perform the full construction of the bulk stress-energy tensor in CLE. In particular, we show the two main properties that characterise the stress-energy tensor: its conformal Ward identities for single insertions into CLE probability functions (Ward identities hold for conserved current associated to symmetries), and its properties under conformal transformations, involving the Schwarzian derivative. We also study the one-point function of the stress-energy tensor, and relate it to what we call the {\em relative partition function} through a certain {\em conformal derivative}. An analysis using CFT arguments shows that the relative partition function is a particular ratio of ordinary partition functions, and that it indeed gives rise to the one-point function.

CLE is a wide generalisation of Schramm-Loewner evolution (SLE), a probabilistic theory for a conformally invariant, fluctuating single curve connecting two boundary points of a domain, introduced in the pioneering work by Schramm \cite{S00} (see the reviews \cite{C05,BB06}). In the context of a particular SLE measure with a property of conformal restriction, the stress-energy tensor was already constructed, first on the boundary \cite{FW02,FW03}, then in the bulk \cite{DRC}. This SLE measure corresponds to a Virasoro central charge equal to 0, and essentially to a CLE where ``no loop remains.'' As was explained in \cite{I}, there is no way of constructing the stress-energy tensor as a local variable in other SLE measures (with non-zero central charge), because one needs to consider all loops, which are not described by SLE. The present work evolved from \cite{DRC}, generalising it to the case of a non-zero central charge. In particular, it is the presence of infinitely many small loops at every point, a property of the CLE measure \cite{W05a}, that provides a central charge.

Some of the techniques used in the present paper for the construction of the bulk stress-energy tensor are in closed relation with those of \cite{DRC}. In particular, the object representing the stress-energy tensor is of similar type to that of \cite{DRC}, and the basic idea behind the derivation of the conformal Ward identities is the same. The main differences, due to the subtleties of CLE, are as follows. First, we introduce the concept of {\em renormalised probability} -- this is the central concept of our construction. It is not a probability in the proper sense, but related to a probability via a certain limit. Conformal invariance of CLE probabilities is lost into a conformal {\em covariance}, but contrary to CLE probabilities, it satisfies a strict {\em conformal restriction} property. The latter is what allows us to use the basic ideas of \cite{DRC}, and the former provides a part of the non-zero central charge. Second, the transformation properties of the stress-energy tensor are derived in a completely different way, in order to take into account the non-zero central charge. These transformation properties constitute the most non-trivial result of this paper. Finally, the one-point function of the stress-energy tensor in CLE needs special care because there are no other fields present, contrary to the SLE case (where there are boundary fields representing the anchoring points of the curve). It is our analysis of the one-point function that led us to define the relative partition function.

The three new objects that we introduce and study are described in definitions \ref{preg}, \ref{T} and \ref{defiZ}. The main results are theorems \ref{theowardplane}, \ref{theoex1pf} and \ref{theowarddom} (conformal Ward identities and one-point function) and theorem \ref{theotransfo} (transformation properties). In the present paper, the only assumption that we must make about CLE is that of differentiability (along with some properties of derivatives), assumption \ref{assdiff}. References to theorems and definitions that are found in the first part of this work \cite{I} will be labelled in the form I.x.x, where x.x is the label used in \cite{I}.

In the present paper, $\hC$ denotes the Riemann sphere $\C \cup \{\infty\}$, and domains are open subsets of $\hC$.

This paper is organised as follows. In section 2, for completeness we review some notions that will be of use here: the main elements of CLE in the dilute regime (we recommend the reader to refer to \cite{I} and to the original works \cite{W05a,ShW07} for more precision), the notion of conformal derivative developed in \cite{diff}, and some elements of conformal field theory. In section 3, for clarity we overview the main constructions and results of the present paper. In section 4, we introduce and study the concept of renormalised probabilities. In section 5, we introduce the CLE definitions of the stress-energy tensor and of the relative partition function, and derive the conformal Ward identities as well as the formula for the one-point function. In section 6, we derive the transformation properties of the CLE stress-energy tensor. In section 7, we suggest the universality of our construction. Finally, in section 8, we present an extensive discussion of our results, making connections with general QFT notions and with standard CFT arguments, providing interpretations for our construction of the stress-energy tensor and for the random loops of CLE, and presenting our perspectives.

{\bf Acknowledgments}

I would like thank D. Bernard, J. Cardy and P. Dorey for illuminating discussions and for suggesting improvements and paths to explore at various stages of this work, and W. Werner for teaching me CLE. I acknowledge the hospitality of the Centre de Recherche Math\'ematique de Montr\'eal (Qu\'ebec, Canada), where part of this work was done and which made many discussions possible (August 2008).

\sect{Preliminaries}\label{sectprel}

\subsection{Conformal loop ensembles}\label{ssectCLE}

As mentioned, this paper is the second part of a work started in \cite{I}, and is based on results obtained there. For completeness, let us recall some of the concepts and objects discussed in \cite{I}, as well as some of the notation introduced.

\begin{figure}
\bc
\includegraphics[width=7cm,height=7cm]{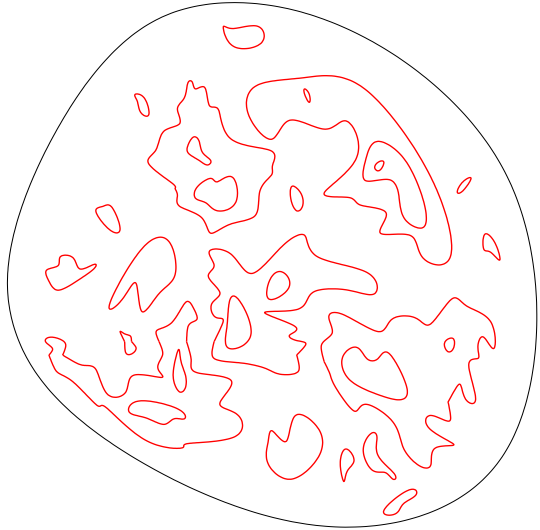}
\ec
\caption{Drawing representing a CLE loop configuration on a domain.}
\label{figCLE}
\end{figure}

Conformal loop ensembles (CLE) are random loop constructions with properties of conformal invariance. The setup to which the present work applies is that of the dilute regime, developed in \cite{W05a,ShW07}; see the first part of this work \cite{I} for an overview of the defining axioms of conformal loop ensembles in this regime, of some of their properties, and of their relation to conformal field theory (CFT). In the dilute regime, any given configuration is composed of a countable infinity of simple loops that do not intersect each other, supported in a simply connected domain (which we will refer to as the domain of definition) -- see figure \ref{figCLE} for a cartoon representation of a configuration.
Conformal loop ensembles provide a measure for each simply connected domain of definition. These measures have properties of conformal invariance: they are invariant under conformal transformations that preserve the domain of definition, and measures on different domains are related to each other by conformal transport. Besides these conformal invariance properties, measures are also related to each other by the nesting and conformal restriction properties. Nesting says that inside any (appropriately) chosen loop $\gamma$, the loops are controlled by the CLE measure in the domain of definition delimited by $\gamma$. Conformal restriction says something similar, but has to do with the outside of loops and of a selected subdomain of the domain of definition. Conformal invariance along with nesting and conformal restriction essentially define conformal loop ensembles. There is a one-parameter family of solutions to these defining conditions. The loops almost surely look locally like SLE$_\kappa$ curves, and one can parametrise the family of CLEs by $\kappa$. In the dilute regime, we have $8/3<\kappa\leq 4$. Conformal loop ensembles are the natural generalisation of Schramm-Loewner evolution, where all loops are being considered in the underlying statistical model.

We will use the symbol $P(\cdot)_C$ for representing the CLE probability function on the domain of definition, or more generally the region of definition, $C$. Although the CLE constructions \cite{W05a,ShW07} only apply to simply connected domains, in \cite{I} we proposed formulae for CLE probabilities on the Riemann sphere $\hC$ and on annular domains, obtained from CLE probabilities on simply connected domains. We showed that these probability functions also satisfy properties of conformal invariance, under certain natural (but non-trivial) assumptions. We will make wide use of such regions of definition below.

Since we are interested in studying probabilities on $\hC$ as well as on domains in $\hC$, all events that we will consider are subsets of the set of configurations of unintersecting loops on the Riemann sphere. When considering probabilities on $C$, we implicitly restrict the event to the set of configurations where all loops are supported on $C$. For an event $\tou$, this restriction is denoted $\tou_C$. Hence, $P(\tou)_C = P(\tou_C)_C$. An important concept introduced in \cite{I} is that of support of an event $\tou$, denoted $\supp(\tou)$. This is a closed set in $\hC$ that essentially tells us in which region the event ``feels'' the loops. See \cite{I} for a more complete description of the CLE events considered, and of the concept of support.

\begin{figure}
\bc
\includegraphics[width=7cm,height=7.6cm]{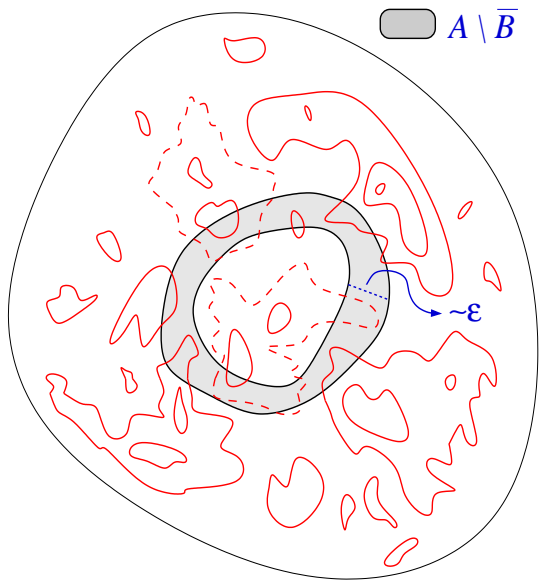}
\ec
\caption{The event $\ev(A,\varep,u)$ (where the partner of $A$ is $B$) on the configuration depicted in figure \ref{figCLE}. The dashed CLE loops break the conditions of the event.}
\label{figevents}
\end{figure}

In the context of the constructions in the first part of this work as well as here, the most useful events are those denoted $\ev(A,\varep,u)$ in \cite{I}. In this notation, $A$ stands for any simply connected domain, $\varep>0$ and $u:\p A\to \hC$ is such that for any $\varep$ small enough, $(\id + \varep u)(\p A) = \p B$ for some simply connected domain $B$ with $\cl{B}\subset A$. That is, the notation implies two simply connected domains $A$ and $B$, whose boundaries are a distance of order $\varep$ away from each other; by convention, we call $B$ the partner of $A$. The event denoted by this symbol is simply that no loop intersect both $\p A$ and $\p B$. See figure \ref{figevents} for a representation.
When $\varep\to0$, this has the effect of ``separating'' the regions $B$ and $\hC\setminus \cl{A}$, so that loops in both regions should become independent of each other. However, this is a very loose description, in particular due to the fact that as $\varep\to0$, the measure of the event tends to zero. Indeed, in CLE, around any point there is almost surely an infinity of loops \cite{W05a}. In \cite{I}, these events are studied at length; in particular, they are used in constructing CLE probabilities on annular domains. They were also observed to enjoy a Lipschitz continuity property, which will provide support to the assumption of differentiability that we will need.

\subsection{Conformal derivatives} \label{ssectdiff}

As we said, our claim that our construction gives the stress-energy tensor in the context of CLE is based on two properties that essentially define it in CFT: the conformal Ward identities that its correlation functions satisfy, and its transformation properties under conformal mappings. The form of the conformal Ward identities that naturally occurs in the context of CLE is different from the one found in standard CFT works, in particular in the case of models with a boundary. In CLE, in order to express the Ward identities in their full generality, we need the concept of conformal differentiability (a particular case of Hadamard differentiability). Conformal derivatives are derivatives, in ``directions'' of small conformal transformations, with respect to sets or objects that are potentially continuous; for instance, the boundary of the domain of definition, or the set upon which some CLE events may naturally depend. This concept is introduced in \cite{diff}, where we also show how it leads to a more compact form of the standard conformal Ward identities of CFT\footnote{Ideas of derivatives with respect to domain boundaries in directions of conformal transformations were also used in \cite{DRC}, although not in relation to the Ward identities, rather in order derive the transformation properties of the stress-energy tensor in the SLE context. However, the concept was not developed to any extent, and there were unfortunately some incorrect statements in intermediate steps.}. Here we review the general theory of conformal derivatives, and in the next subsection, we explain how conformal derivatives are involved in the conformal Ward identities. Such derivatives are also involved in the expression for the central charge and for the one-point function of the stress-energy tensor obtained in the present work.

Suppose that we have a space with, at a point, a well-defined action of transformations conformal on a domain $A$ and near enough to the identity: for instance, the space of closed subsets of $A$ with action by conformal mapping of subsets (and any point in that space), or the space of conformal transformations on $A$ with action by composition, right or left (and, again, any point in that space). The set of all conformal transformations near to the identity (in an appropriate sense) defines a neighbourhood of this point, and the associated algebra spans the tangent linear space. Then, we can roughly define $A$-differentiability \cite{diff} at this point by the condition that a function ($\R$-valued or $\C$-valued, or valued in some normed linear space), defined on a neighbourhood of this point, change by a small amount under conformal transformations that are small on $A$. In the present work, we will only need the cases where $A$ is a simply connected domain.

Note that in CLE and in CFT, we are often working with objects (probability functions or correlation functions) that are invariant or covariant under conformal transformations. Hence, it may seem {\sl a priori} that derivatives along conformal transformations should be somewhat trivial -- it is in the moduli space that we should differentiate in order to get non-trivial variations. But recall that conformal invariance or covariance only holds for very particular sets of conformal transformations. For instance, on the Riemann sphere, only global conformal transformations lead to invariance or covariance, and on a domain $C$, only transformations that are conformal on $C$ do so. As we will see in the next subsection, it is when looking at derivatives along other conformal transformations that we obtain interesting results; these other conformal transformations indeed change the moduli.

Let us recall the main results \cite{diff}. Consider $A$ a simply connected domain that does not contain $\infty$. Consider a family of transformations $\{g_\eta,\,\eta>0\}$ that are conformal on $A$ for all $\eta$ small enough, and write $g_\eta = \id + \eta h_\eta$. Suppose that $h_\eta$ converges uniformly on any compact subset of $A$ as $\eta\to0$ to a function $h$. Note that $h$ is holomorphic on $A$. Then $A$-differentiability of a function $f$ at the point $\Sigma$ implies that there exists, uniquely, two functions $\Delta_{a;z}^A f(\Sigma)$ and $\b\Delta_{a;\b{z}}^A f(\Sigma)$ depending on a parameter $a\in\hC\setminus A$, holomorphic and antiholomorphic, respectively, outside $A$ as functions of $z$ and zero at $z=a$, such that
\beq\label{derf}
	\nabla_h f(\Sigma) := \lim_{\eta\to0} \frc{f(g_\eta(\Sigma)) - f(\Sigma)}{\eta} =
		\int_{z\in \vec\p A^-} \dd z \,h(z) \Delta_{a;z}^A f(\Sigma) +
			\int_{z\in \vec\p A^-} \bd \b{z} \,\b{h}(\b{z}) \b\Delta_{a;\b{z}}^A f(\Sigma).
\eeq
Here, we define for convenience
\beq\label{ddz}
	\dd z = \frc{dz}{2\pi i},\quad \bd \b{z} = -\frc{d\b{z}}{2\pi i}.
\eeq
The notation $\vec\p A$ means the {\em oriented} boundary of $A$, indicating that the contour of integration is in the counter-clockwise direction around the interior of $A$. Also, the superscript $^-$ in $\vec\p A^-$ indicates that the contour is on a path inside the domain $A$ but infinitesimally close to its boundary $\p A$. The functions $\Delta_{a;z}^A f(\Sigma)$ and $\b\Delta_{a;\b{z}}^A f(\Sigma)$ are simply complex conjugate of each other, so it is sufficient to discuss the holomorphic part.

The same equation holds if $\infty\in A$, where $h$ is holomorphic on $A$ except possibly for a pole of order 2 at $z=\infty$ (i.e.\ behaves as $O(z^2)$). More precisely, in this case, the set of all families $\{g_\eta,\,\eta>0\}$ for which the limit in (\ref{derf}) is required to exist is simply obtained by conformal transport from a domain excluding $\infty$. Also, in this case, the unique function $\Delta_{a;z}^A f(\Sigma)$ is required to be holomorphic in $A$ except for a pole of order no more than 3 at $z=a$.

We call $\nabla_h g(\Sigma)$ the {\em conformal derivative of $f$ at $\Sigma$ in the direction $h$}. It is shown in \cite{diff} that the limit that gives its definition in (\ref{derf}) only depends on $h$, no matter what precise family $\{g_\eta,\,\eta>0\}$ we take (that is, what domain $A$ we choose).

Naturally, equation (\ref{derf}) by itself does not uniquely define the functions $\Delta_{a;z}^A f(\Sigma)$ and $\b\Delta_{a;\b{z}}^A f(\Sigma)$ involved: there are two classes of functions, the {\em holomorphic and antiholomorphic $A$-classes}, that could be used. These classes are completely characterised by the singularity structure in $A$ of the functions they contain (and are naturally related to the Hadamard derivative when we see conformal differentiability in the context of Hadamard differentiability). But we choose the particular members $\Delta_{a;z}^A f(\Sigma)$ and $\b\Delta_{a;\b{z}}^A f(\Sigma)$ of the classes, with the additional requirements as described above (these requirements uniquely define these particular members). These are called the {\em holomorphic and antiholomorphic $A$-derivatives} (of $f$ at $\Sigma$).

$A$-differentiability of $f$ at $\Sigma$ implies $B$-differentiability of $f$ at $\Sigma$ for any simply connected domain $B$ such that $A\subseteq B$. Also, if $f$ is both $A$- and $B$-differentiable for two domains $A$ and $B$ such that there exteriors have non-zero intersection, then with $a\in \hC\setminus (A\cup B)$, there is a simple map that allows us to obtain $\Delta_{a;z}^A f(\Sigma)$ from $\Delta_{a;z}^B f(\Sigma)$ (see \cite{diff}). In particular, the singularity structure in the domain is preserved. This means that the set of all domains $A$ for which we have $A$-differentiability of $f$ at $\Sigma$ can be divided into partitions: in any given partition, the singularity structure of holomorphic $A$-classes is the same. We will characterise a partition by any one of its member; for instance, the partition that contains a domain $A$ will be called the {\em $A$-partition}. Note that there is at most one partition that contains at least one member $A$ such that $\infty\not\in A$. This will be called the {\em bounded partition}. The set of points that are in all domains $A$ for which we have $A$-differentiability is the {\em fundamental set}. Each connected component of the complement of this set on $\hC$ corresponds to a distinct partition. These components constitute the {\em holomorphy regions} of the various partitions, where the holomorphic derivatives are holomorphic functions (except possibly of a pole of order 3 at $\infty$ as explained above).

For instance, if we are looking at a function of sets, in a neighbourhood of a set $\p C$ that is the boundary of a simply connected domain $C$, then we may expect to have $A$-differentiability at $\Sigma = \p C$ for any simply connected domain $A$ that contains $\p C$. In this case, we would have two partitions, the $A$-partition, containing all $A$ such that $\cl{C}\subset A$, and the $B$-partition containing all $B$ such that $\hC\setminus C \subset B$. Hence, in this case we would have only two essentially different holomorphic classes, or holomorphic derivatives. The two holomorphy regions would simply be the two domains delimited by $\p C$.

There is an important situation where many simplifications occur: when the function $f$ is invariant, at $\Sigma$, under global conformal transformations $G$ in a neighbourhood of the identity: $f(G(\Sigma)) = f(\Sigma)$. Then, we can define the {\em global holomorphic $A$-derivative} (of $f$ at $\Sigma$): if $\infty\not\in A$, it is defined by $\Delta_z^A f(\Sigma) = \Delta_{\infty;z}^A f(\Sigma)$, and if $\infty\in A$, it is defined by $\Delta_z^A f(\Sigma) = \Delta_{a;z}^A f(\Sigma)$ for any $a\in\hC\setminus A$. It turns out \cite{diff} that this is an unambiguous definition, and that for any $A$ and $B$ (containing or not $\infty$) in a given partition, we have $\Delta_z^A f(\Sigma) = \Delta_z^B f(\Sigma)$. Hence, we need to keep $A$ in the notation $\Delta_z^A f(\Sigma)$ for the sole purpose of identifying the partition, if there are many partitions. The global holomorphic derivative has the properties that it is exactly holomorphic on the holomorphy region of the partition, and that for the bounded partition, it behaves like $O(z^{-4})$ as $z\to\infty$. In the latter case, the coefficient of $z^{-4}$ is proportional to what we call the {\em (holomorphic) charge of $f$ at $\Sigma$}, denoted $\chrg f(\Sigma)$. More precisely,
\beq \label{cancharge}
	\Delta_z^A f(\Sigma) = \frc{\chrg f(\Sigma)}{32} z^{-4} + O(z^{-5}) \quad \mbox{for $A$ in the bounded partition}.
\eeq
This coefficient, for appropriate $f$ and $\Sigma$, is what is related to the central charge in our construction of the stress-energy tensor. The antiholomorphic charge, $\b\chrg f(\Sigma)$, is likewise defined from the global antiholomorphic derivative.

The global holomorphic derivative has a very simple transformation property under global conformal transformations $G$ (not necessarily near to the identity):
\beq \label{cantrans}
	\Delta_z^A (f\circ G)(\Sigma) = (\p G(z))^2 \Delta_{G(z)}^{G(A)} f(G(\Sigma)).
\eeq
In fact, another important simplification occurs when $f$ is invariant, at $\Sigma'$, under transformations conformal on a simply connected domain $A$ in a neighbourhood of the identity. In this case, let us consider the global holomorphic $B$-derivative for a simply connected domain $B$ such that $\hC\setminus B \subset A$. Then, under appropriate continuity conditions \cite{diff}, we have a transformation property as in (\ref{cantrans}), but for any $g$ conformal on $A$ (and also not necessarily near to the identity)
\beq \label{cantrans2}
	\Delta_z^B (f\circ g)(\Sigma) = (\p g(z))^2 \Delta_{g(z)}^{\hC\setminus g(\hC\setminus B)} f(g(\Sigma))
\eeq
where $g(\Sigma) = \Sigma'$. Of course, if $A$ and $B$ turn out to be in the same partition, then both sides are exactly zero: invariance under transformations conformal on $A$ imply that the holomorphic $A$-derivative vanishes, hence all holomorphic derivatives vanish in the $A$-partition. But this formula is non-trivial when $A$ and $B$ are in different partitions (and if there is no invariance under transformations conformal on $B$); this can well be the case since $\hC\setminus B \subset A$, and this is what occurs in the applications that interest us here.

Naturally, the usual chain rule of calculus holds here as well. Let us consider for simplicity the case where the function $f$ differentiated is valued in $\R$ -- this is the only case that we need. Then, if $F$ is a function on $\R$ differentiable at the point $f(\Sigma)$, we have
\beq\label{chainrule}
	\Delta_{a;z}^A (F\circ f)(\Sigma) = \Delta_{a;z}^A f(\Sigma) \, \lt.\frc{dF(t)}{dt}\rt|_{t=f(\Sigma)}.
\eeq

Note that upon considering the analytic structure of $h(z)$ and of $\Delta_{a;z}^A f(\Sigma)$ in (\ref{derf}), it is usually possible to omit the superscript $^-$ in the condition $z\in \vec\p A^-$ determining the integration path. Indeed, we can often deform the path all the way to $\vec\p A$ without problem. In most situations that occur in the present work, this is the case.

The domains $A$ that we will consider will often be of the form $\hC\setminus \cl{N(w)}$ where $N(w)$ is the neighbourhood of some point $w$, that can be infinity. The choice of the neighbourhood will not affect the results (as long as it satisfies certain conditions as stated when required). We will denote
\beq \label{Cw}
	\hC_w = \hC\setminus \cl{N(w)}.
\eeq

When there may be confusion, we will indicate by a subscript $_{|\,\Sigma}$ the argument with respect to which the differentiation occurs. For instance, we write $\Delta_{z\,|\,\Sigma}^A F(f(\Sigma)) = \Delta_{z}^A (F\circ f)(\Sigma)$, and similarly for $\chrg_{|\,\Sigma} F(f(\Sigma))$.

\subsection{Conformal field theory}\label{ssectCFT}

We now provide a description of the basic structure of conformal field theory, purely from the viewpoint of correlation functions and their properties under conformal transformations.

A conformal field theory model, in a basic description, can be seen as follows. It is 1) a region of definition $C$, for us this can be a domain in $\hC$ or the Riemann sphere $\hC$ itself, 2) an infinite-dimensional vector space over some field of complex functions on $C$ (the vector space of local fields), and 3) a family of multilinear maps from the $n^{\rm th}$ tensor power of this vector space to some space of complex functions of $n$ non-coincident points on $C$ (the correlation functions), for $n=1,2,\ldots$ (the number of fields in the correlation functions). The points are understood as the positions of the local fields involved in the correlation functions. Denoting a discrete set of $n$ local fields by $\Or_1,\,\Or_2,\,\ldots,\,\Or_n$, and the region of definition by $C$, the correlation function evaluated at the positions $z_1,\,z_2,\,\ldots,\,z_n$ is denoted by
\beq
	\bra \Or_1(z_1) \Or_2(z_2)\cdots\Or_n(z_n)\ket_C.
\eeq
In general, if $\Or$ is a field, then $\p \Or$ and $\b\p \Or$ also are fields, whose correlation functions are the holomorphic and antiholomorphic derivatives, with respect to the position, of those of $\Or$.

This structure is what occurs naturally when considering a CFT model as the scaling limit of a lattice statistical model at a critical point. Intuitively, through the scaling limit, every field in a correlation function corresponds to a statistical variable at some lattice position, and the correlation function itself corresponds to the average of the product of these local statistical variables at various positions. The scaling limit of a statistical model at a critical point is obtained, roughly speaking, by sending the lattice spacing to zero (in other words, by making the lattice positions of the variables very far apart on the lattice), while ``renormalising'' the statistical variables in such a way that the average converges. The result is expected to be a CFT correlation function, where the various proportions of the positions of the fields are in agreement with the fixed proportions taken by the positions of the statistical variables in the scaling process. The renormalisation requirement means that we must in fact choose, instead of a statistical variable at a lattice position, an appropriate finite {\em linear combination} of statistical variables, all at or near to a given position, with coefficients that depend on the lattice spacing. In general, these coefficients must diverge as the lattice spacing is sent to zero, in such a way that the resulting average has a finite limit. The space of correct linear combinations of statistical variables is expected to be the space of local fields of the CFT.

Correlation functions in CFT are expected to satisfy a wealth of properties as functions of the positions of the fields. One of them is conformal invariance (or covariance): there exists an automorphism of the vector space of local fields that is equivalent, from the viewpoint of correlation functions, to a change of field positions and of the region of definition by a conformal map. For a conformal transformation $g:C\to C'$, this can be written as:
\beq\label{invCFT}
	(g\cdot\Or_j)(g(z)) = \sum_i q_{j,i}(g;z)\Or_j^{(i)}(g(z)),\quad
	\bra \prod_{j=1}^n (g\cdot\Or_j)(g(z_j))\ket_{g(C)} = \bra \prod_{j=1}^n \Or_j(z_j)\ket_C
\eeq
for some (not necessarily holomorphic) functions $q_{j,i}(g;\cdot)$ defined on $C$. These functions determine the transformation property of the field $\Or_j$ under conformal mappings, and are assumed to be independent of $C$ by locality. When $C=C'$, this is a symmetry, or an invariance, of the conformal field theory on $C$, and when $C\neq C'$ we talk about conformal transport.

Naturally, the automorphisms must agree with the group of conformal transformations. A possibility is
\beq\label{primary}
	(g\cdot\Or)(g(z)) = (\p g(z))^{\delta} (\b\p \b{g}(\b{z}))^{\t{\delta}} \Or(g(z))
\eeq
for some real $\delta$, $\t{\delta}$, the holomorphic and antiholomorphic conformal dimensions of the field. The scaling dimension is $d=\delta+\t{\delta}$ and the spin is $s=\delta-\t{\delta}$; these describe how the field transforms under scaling transformations and under rotations. A field with transformation property (\ref{primary}) is called a {\em primary field} \cite{BPZ}, and will be said to have dimension $(\delta,\t\delta)$. It is generally assumed in (rational) CFT that there is a finite-dimensional subspace of primary fields. The main idea behind primary fields is that they only ``feel'' the conformal transformation locally: they arise from statistical variables that are ``local enough.''

When there is a symmetry or invariance in a local QFT model, there are associated Noether current and conserved charge generating the transformation upon which there is invariance. The invariance equation then follows, in a Hilbert space formulation, from the fact that the conserved charge commutes with the Hamiltonian and annihilates the ground state (the conservation conditions for the charge, consequence of the conservation of the Noether current). In this context, as a consequence of conformal invariance in CFT, one expects that there are particular fields, forming the {\em stress-energy tensor}, $T(w)$ and $\b{T}(\b{w})$, whose correlation functions have the following properties \cite{F84,BPZ} (for tutorials, see, for instance, \cite{Gins,DFMS97}): 1) they are, respectively, {\em holomorphic} and {\em antiholomorphic} in $w$ on the domain of definition (whence the choice of arguments) except at the positions of other local fields; and 2) for any conformal transformation of the form $g = \id + \eta h$ with $g(C)=C$ (so that we are talking about a true symmetry) that is near to the identity $\id$ (that is, with $\eta>0$ small), one expects that
\beqa\label{conftrCFT}
	\lefteqn{\bra (g\cdot\Or_1)(g(z_1)) \prod_{j=2}^n \Or_j(z_j) \ket_{C}} && \\
	&=& \bra \prod_{j=1}^n \Or_j(z_j) \cdots \ket_{C}
	+ \eta 
	\oint_{z_1} \bra \lt[\dd w\, h(w) T(w) + \bd\b{w}\, \b{h}(\b{w}) \b{T}(\b{w})\rt] \prod_{j=1}^n \Or_j(z_j)\ket_{C}
	+ o(\eta) \no
\eeqa
where the contour of integration is around the point $z_1$ and chosen in such a way that it is continuously deformable to $z_1$ without crossing any other field positions. The fact that the contour can be deformed without changing the result is just from the holomorphy/antiholomorphy of the correlation functions as functions of $w$, and expresses the conservation of the current. The fact that the transformed field can be written, for $\eta$ small, through the integration of a current is the expression that the associated charge generates the transformation. Then, the covariance equation (\ref{invCFT}), in infinitesimal form and in the case $g(C)=C$, is simply
\beq\label{invTCFT}
	\oint_{z_1,\ldots,z_n}
	\bra \lt[\dd w\, h(w) T(w) + \bd\b{w}\, \b{h}(\b{w}) \b{T}(\b{w})\rt] \prod_{j=1}^n \Or_j(z_j) \ket_{C} = 0
\eeq
where the integration contour surrounds all field positions.

Another popular way of viewing the origin of the stress-energy tensor, in a sense more natural in the context of statistical models, is through its involvement in describing the effect of metric changes \cite{F84}; see section \ref{discuss} where this idea is used.

Let us now consider the product $T(w) \Or(z)$ inside correlation functions, as a function of $w$. Expanding it in a power series in $w-z$ with coefficients that are other local fields at $z$, this is Wilson's operator product expansion. One of the main expected properties of local QFT is that this expansion is independent from the other fields inside the correlation functions, and from the domain of definition. From this, the conditions (\ref{conftrCFT}) imply the following operator product expansion for the stress-energy tensor with a primary field:
\beq\label{OPET}
	T(w) \Or(z) = \frc{\delta}{(w-z)^2}\Or(z) + \frc{1}{w-z} \frc{\p}{\p z} \Or(z) + \mbox{regular terms in $w-z$},
\eeq
which is to be understood as valid inside any correlation functions. This is an extremely strong condition. In particular, it is consistent with $T(w)$ transforming, under rotations and scaling, like (\ref{primary}) with $\delta=2,\,\t{\delta}=0$ (i.e.\ with scaling dimension 2 and spin 2). On $\hC$, we may look at the situation where $w$ is very far from all other fields. By factorisation of local QFT and since the stress-energy tensor has zero average on $\hC$ by rotation covariance, this limit is simply 0. Along with the operator product expansion, this completely fixes the dependence on $w$ of the correlation function \cite{BPZ}:
\beq\label{wardidCFTplane}
	\bra T(w) \prod_{j=1}^n \Or_j(z_j) \ket_\hC =
	\sum_{j=1}^n \lt(\frc{\delta_j}{(w-z_j)^2} + \frc1{w-z_j} \frc{\p}{\p z_j} \rt) \bra \prod_{j=1}^n \Or_j(z_j) \ket_\hC.
\eeq
In other words, the operator product expansion determines the singularity structure in $w$, and the factorisation property tells us that as $w\to\infty$, the function should vanish; the above is the only solution to this simple ``Riemann-Hilbert problem.'' This is what we refer to as the conformal Ward identity on $\hC$. Here, it is obtained for primary fields, but any other field, with known transformation property, can be dealt with in similar ways \cite{BPZ}: its transformation property fixes the singularity structure in $w$ at its position.

One of the main consequences of the algebraic analysis of the operator product expansion is the transformation property of the stress-energy tensor itself \cite{BPZ}. It turns out that it may transform ``anomalously''; that is, although (\ref{OPET}) is consistent with $T(w)$ having, under rotations and scaling transformations, a scaling dimension 2 and a spin 2, there may be an extra term to (\ref{primary}) under other conformal transformation. The anomalous term is associated to the unique central extension of the Witt algebra (the algebra of infinitesimal conformal transformations), the Virasoro algebra. The stress-energy tensor is expected to transform as
\beq\label{transTCFT}
	(g\cdot T)(g(w)) = (\p g(w))^2 T(g(w)) + \frc{c}{12} \{g,w\}
\eeq
where $\{g,w\}$ is the Schwarzian derivative:
\beq
	\{g,w\} = \frc{\p^3 g(w)}{\p g(w)} - \frc32 \lt(\frc{\p^2g(w)}{\p g(w)}\rt)^2.
\eeq
The constant $c$ is the Virasoro central charge, and is a characteristic of the CFT model under study.

Then, on any simply connected domain in $\hC$, it is also possible to completely fix the dependence on $w$. Let us consider the upper-half plane $\uH$. There, $\bra T(w)\ket_\uH = 0$ by, for instance, transport from the unit disk, and since the average is zero on the unit disk by rotation covariance. On $\uH$, the invariance condition (\ref{invTCFT}) is in agreement with, although does not immediately imply, the local condition $T=\b{T}$ on $\R$. In CFT this local condition is simply assumed to hold \cite{C84}, based on QFT arguments. From this, and from analyticity and factorisation considerations as in the case of $\hC$, the conformal Ward identity on $\uH$ is
\beq\label{wardidCFTH}
	\bra T(w) \prod_{j=1}^n \Or_j(z_j)\ket_\uH = \sum_{j=1}^n \lt(\frc{\delta_j}{(w-z_j)^2} + \frc1{w-z_j} \frc{\p}{\p z_j} +
	\frc{\t{\delta}_j}{(w-\b{z}_j)^2} + \frc1{w-\b{z}_j} \frc{\p}{\p \b{z}_j} \rt) \bra \prod_{j=1}^n \Or_j(z_j) \ket_\uH.
\eeq
By conformal transport using (\ref{transTCFT}), we may then obtain similar identities on any simply connected domain, determining the full $w$ dependence.

From all these consideration, we obtain expressions for the connected correlation functions
\beq\label{wardidC}
	\bra T(w) \prod_{j=1}^n \Or_j(z_j)\ket^{(c)}_C :=
	\bra T(w) \prod_{j=1}^n \Or_j(z_j) \ket_C - \bra T(w) \ket_C \bra \prod_{j=1}^n \Or_j(z_j) \ket_C
\eeq
in terms of differential operators on $\bra \prod_{j=1}^n \Or_j(z_j) \ket_C$ for any simply connected domain $C$ and for $C=\hC$. These expressions do not involve the central charge, thanks to the subtraction of the disconnected term. In particular, these connected correlation functions vanish as $w\to\infty$, and transform as if $T(w)$ were a primary field of dimension $(2,0)$. The relations we obtain for connected correlation functions are what we will call the {\em conformal Ward identities on $C$}. Note that for $C=\uH$ or $C=\hC$, for instance, the connected correlation functions of the stress-energy tensor are equal to the correlation functions themselves.

The conformal Ward identities and the transformation properties of the stress-energy tensor are its two main properties. As we mentioned above, one goal of this paper is to recover these in the CLE context, thus providing a more mathematically satisfying way than that outlined above using CFT and more general QFT principles. For this, we need to recast the conformal Ward identities in a form involving conformal derivatives \cite{diff}.

A comparison of (\ref{invTCFT}) and (\ref{derf}) suggests that the holomorphic and antiholomorphic $A$-derivatives should be related to the holomorphic and antiholomorphic stress-energy tensor components $T(w)$ and $\b{T}(\b{w})$, for some $A$. However, the relation is not direct. Let us consider the function $f$, on a space of conformal transformations, defined by
\beq
	f(g) = \bra \prod_{j=1}^n (g\cdot \Or_j)(z_j)\ket_{[g(\p C)]}
\eeq
for any transformation $g$ conformal on a simply connected domain containing the set $\{z_1,\ldots,z_n\}\cup \p C$. Here, $[g(\p C)]$ denotes the domain bounded by $g(\p C)$ and containing the points $g(z_1),\ldots,g(z_n)$; note that $g$ does not need to be conformal on $C$. If $C=\hC$, then we simply take $\p C = \emptyset$. Naturally, by conformal invariance or conformal transport, we have that $f(g) = f(\id)$ for any $g$ that is conformal on $C$. At the point $g$ on the space where $f$ is defined, there is a well-defined action of transformations $g'$ conformal on $\{g(z_1),\ldots,g(z_n)\}\cup g(\p C)$: the result is simply $g'\circ g$. Under this action, $f$ is invariant under global conformal transformations for any $C$, including $C=\hC$. Hence, $f$ has a well-defined global holomorphic derivative.

It is proven in \cite{diff} that the conformal Ward identities for the connected correlation functions (\ref{wardidC}) are equivalent to the identification
\beq\label{wardgen}
	\bra T(w) \prod_{j=1}^n \Or_j(z_j) \ket_C^{(c)} =
		\Delta_w^{\hC_w} f(\id)
\eeq
(for $w\in C$). See (\ref{Cw}) for the notation $\hC_w$; here, the neighbourhood $N(w)$ does not intersect $\{z_1,z_2,\ldots,z_n\}\cup \p C$. That is, the insertion of the (connected part of the) holomorphic stress-energy tensor at $w$ is obtained by taking the global holomorphic $\hC_w$-derivative evaluated at $w$.  Using the function
\[
	h_{w,\theta}(z) = \frc{e^{i\theta}}{w-z}
\]
and from holomorphy of the global holomorphic derivative, this can be written as well in the forms
\beqa
	\bra T(w) \prod_{j=1}^n \Or_j(z_j)\ket_C^{(c)}
\label{wardgen2}		
	 &=&
		\int_{z\in\vec\p \hC_w} \frc{\dd z}{w-z} \Delta_z^{\hC_w} f(\id) \\
	&=&
		\int_0^{2\pi} \frc{d\theta}{2\pi} e^{-i\theta} \,\nabla_{h_{w,\theta}} f(\id).
\label{wardgen3}
\eeqa
Relations (\ref{wardgen}), (\ref{wardgen2}) and (\ref{wardgen3}) are proven to hold in \cite{diff} for any simply connected domain $C$ and for $C=\hC$, and are expected to hold in general.

If the fields involved are primary fields of dimension $(0,\t{h})$, then we can consider more simply $f$ as a function on the space of sets of the form $\{z_1,\ldots,z_n\}\cup \p C$, with the natural action of conformal mappings on these sets. The same formulae hold, so that we can write, for instance
\beq\label{wardzero}
	\bra T(w) \prod_{j=1}^n \Or_j(z_j) \ket_C^{(c)} =
		\Delta_{w}^{\hC_w} \bra \prod_{j=1}^n \Or_j(z_j) \ket_C
\eeq
where here $\Delta_{w}^{\hC_w} \cdots \equiv \Delta_{w\,|\,\{z_1,\ldots,z_n\}\cup \p C}^{\hC_w}\cdots$.

For $C=\hC$, the form (\ref{wardgen3}) immediately leads to the usual conformal Ward identities, in particular to (\ref{wardidCFTplane}) in the case of primary fields. Indeed, we just have to use the basic limit definition of the conformal derivative $\nabla_{h_{w,\theta}} f(\id)$, (\ref{derf}), and evaluate the limit using the primary-field transformation properties (\ref{primary}). Hence in this case, the result (\ref{wardgen}) is just a simple re-writing of the usual conformal Ward identities.

In the case where $C$ is a simply connected domain, however, the formula (\ref{wardgen}) is non-trivial. In this case, there are two partitions associated to the conformal derivative, characterised by the domains $\hC_a$ for $a\in \hC\setminus\cl{C}$, and $\hC_w$ for $w\in C$, where again the neighbourhoods do not intersect $\{z_1,z_2,\ldots,z_n\}\cup \p C$. The $\hC_a$-partition is trivial, in the sense that $\Delta_z^{\hC_a} f(\id) = 0$, because $f(g) = f(\id)$ for $g$ conformal on $\hC_a$ by conformal transport or conformal invariance as explained above. However, the $\hC_w$ partition is non-trivial, since there is no conformal invariance for $g$ conformal on $\hC_w$, except if it is a global conformal transformation; it is this non-trivial partition that leads to the stress-energy tensor. Note that we can separate the part of the derivative that applies to the fields from the part that applies to the domain boundary. The part that applies to the fields gives terms similar to those appearing in the case $C=\hC$; the other part gives an extra contribution. For instance, for primary fields (see (\ref{wardidCFTplane}) for the contribution coming from the fields), we have
\beqa
	\lefteqn{\bra T(w) \prod_{j=1}^{n} \Or_j(z_j) \ket_C^{(c)}} && \n &=& \lt[\sum_{j=1}^n
		\lt(\frc{\delta_j}{(w-z_j)^2} + \frc1{w-z_j} \frc{\p}{\p z_j}\rt) +
		\int_{z\in\vec\p (\hC\setminus C)^+} \frc{\dd z}{w-z} \Delta_{z\,|\,\p C}^{\hC_w}\rt]
			\bra \prod_{j=1}^n \Or_j(z_j) \cdots\ket_C. \no
\eeqa
In the last term, the derivative is with respect to $\p C$, and we have moved the contour of integration infinitesimally close to $\p C$, keeping it inside $C$  (this is the meaning of the superscript $^+$). The last term can naturally be interpreted as a contribution from a continuum of zero-dimensional fields forming the boundary of the domain of definition. Likewise, the last term can be interpreted as the analytic behaviour necessary to reproduce the transformation of the domain under conformal transport. Here, the transformation of the domain would be obtained via a formula similar to (\ref{conftrCFT}), but in the case where $g(C)\neq C$, and seeing the boundary $\p C$ as if it were a ``primary field at $\infty$'' of dimension (0,0) (this corresponds to applying the charge associated to the infinitesimal transformation, to the state associated to $\p C$.)

Note finally that expression (\ref{wardgen}) along with (\ref{cantrans2}) is in agreement with the transformation properties of the stress-energy tensor. We will discuss in section \ref{discuss} how the global holomorphic derivative $\Delta_w^{\hC_w}$ comes out also from considering the stress-energy tensor in relation to metric variations, in particular for an expression of the one-point function of the stress energy tensor in terms of partition functions.

\sect{Overview of results}\label{sectoverview}

Consider the events $\ev(A,\varep,u)$ reviewed in subsection \ref{ssectCLE} (and introduced in \cite{I}). Roughly, by taking $\varep\to0$, they allow us to separate the domain $A$ from the domain $\hC\setminus \cl{A}$, in the sense that the CLE random loops in these two domains become independent, since no loop is allowed to intersect $\p A$. In order to take the limit $\varep\to0$, we must normalise the probability: the event has measure zero in this limit because of the presence of the infinity of small loops. With $C$ a simply connected domain or $C=\hC$, we consider:
\[
	\lim_{\varep\to0} \frc{P(\tou,\ev(A,\varep,u))_C}{z(\varep)},
\]
where $z(\varep)$ is an appropriate normalisation that vanishes as $\varep\to0$, in order that the result be finite, and $\tou$ is an event supported in $C$ away from $\p A$. Then, we expect the result to be described by a probability theory where loops are in two disjoint domains, $A$ and $C\setminus \cl{A}$. If $z(\varep)$ is chosen to be $P(\ev(A,\varep,u))_C$, then it was proven in \cite{I} that the result is a CLE probability function on $A$ if $\supp(\tou)\subset A$, or on $C\setminus\cl{A}$ if $\supp(\tou)\subset C\setminus \cl{A}$ (theorems I.5.1, I.5.2, I.5.3 and definition I.5.1). Based on theorem I.5.5, in the next section we will show that at least with an appropriate choice of $u$, that depends on $A$, the normalisation $z(\varep)$ can be chosen independently from both $\tou$ and $A$ (we will take the case where $\supp(\tou)\subset C\setminus \cl{A}$ throughout, which will be sufficient for our purposes). It is likely that there are many possible choices of $u$ that would make this possible, but we will choose certain particular functions $u_A$. The result of the limit, with $u=u_A$ and with appropriate $z(\varep)$ independent of $\tou$ and $A$, is what we call a {\em renormalised probability}, denoted $\Preg(\tou;A)_C$, or $\Preg(A)_C$ if $\tou$ is the trivial event (see definition \ref{preg}). The renormalised probability $\Preg(\tou;A)_C$ should be understood as an appropriately finitised ``probability'', on $C$, of the event $\tou$ in conjunction with the event that no loop intersect the boundary $\p A$ of the region $A$; although it is not a proper probability in that it is not bounded by 1.

Contrary to usual CLE probabilities, renormalised probabilities are not conformally invariant: a conformal transformation $g$ affects $u_A$ to give a function that is not necessarily $u_{g(A)}$, hence the result of the limit is in general different. They are, instead, conformally covariant (theorem \ref{theotransreg}). However, the particular choice $u_A$ that we took (subsection \ref{ssectchoice}) ensures that they are {\em invariant under global conformal transformations} (theorem \ref{invglob}). This invariance is what guided the choice of $u$. Additionally, also contrary to CLE probabilities, renormalised probabilities give rise to {\em exact conformal restriction}: the ratio $\Preg(\tou;A)_C/\Preg(A)_C$ is the probability $P(\tou)_{A\setminus \cl{C}}$ (theorem \ref{theorestreg}). Conformal restriction is the main reason for introducing renormalised probabilities.

\begin{figure}
\bc
\includegraphics[width=5cm,height=5cm]{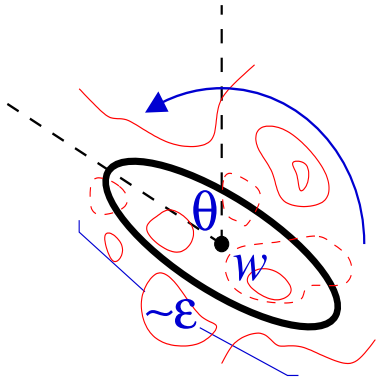}
\ec
\caption{A representation of the process by which the stress-energy tensor is ``inserted'' into a probability function at the point $w$. The thickness of the ellipse (here the thick black curve) centered at $w$ is sent to zero first, in order to obtain a renormalised probability. Thin red curves are CLE loops or arcs thereof, full if they are allowed by the conditions of the renormalised probability, dashed if they break the conditions. Then, a Fourier transform in $\theta$ is taken (corresponding to a rotation with spin 2), and the appropriately normalised limit of a small ellipse $\ep\to0$ is evaluated.}
\label{figT}
\end{figure}

The construction of the stress-energy tensor from renormalised CLE probabilities then follows very closely the construction of \cite{DRC} from ordinary SLE$_{8/3}$ probabilities. In the context of SLE, the event that the curve does not intersect a given region boundary is generally of non-zero measure, so in this context, we directly used probabilities instead of renormalised probabilities. The main ingredient in this construction is conformal restriction, which is a property of both the SLE$_{8/3}$ measure and of renormalised probabilities in CLE. Let us choose $A$ (the domain whose boundary is required not to be intersected) to be a small elliptical domain centered at $w$, of length of order $\ep$, and at an angle $\theta$ with respect to some fixed axis. Taking the second Fourier coefficient with respect to $\theta$ of the renormalised probability $P(\tou;A)_C$, multiplying by $\ep^{-2}$ and by a fixed normalisation constant, and then taking the limit $\ep\to0$, the result is interpreted as the insertion of the stress-energy tensor at $w$ in the probability of the event $\tou$ (see definition \ref{T}). See figure \ref{figT} for a representation of the process.
What we obtain is of course not a probability; it is a limit (over $\ep$) of a linear combination (due to the Fourier transform) of renormalised probabilities. We will refer to this as a {\em pseudo-probability} -- it is more closely related to correlation functions of CFT (see section \ref{sectuniv}).

In \cite{DRC}, it was shown, in the context of SLE$_{8/3}$ on the upper half-plane $\uH$, that the result of this insertion is described by the standard conformal Ward identities on that domain (\ref{wardidCFTH}), and that the resulting object at $w$ transforms like a primary field of dimension $(2,0)$. Since it is known that the central charge of the CFT corresponding to SLE$_{8/3}$ is zero, this was the basis for the identification of this object with the holomorphic stress-energy tensor. In \cite{DRC}, the event $\tou$ was that the SLE curve winds around a set of points in $\uH$, and the Ward identities obtained identified these as zero-dimensional primary fields. The Ward identities also identified the end-points of the curve as boundary primary fields with the correct expected dimension. In fact, the Ward identities were obtained more generally for correlation functions containing many insertion of the stress-energy tensor.

In the present work, we keep the event $\tou$ arbitrary, except that we require that it depends on a set $\Sigma$ contained by $\supp(\tou)$ in such a way that the action of a conformal transformation on $\tou$ is reproduced by the action on $\Sigma$ (this is not essential, but applies to most events of interest, and simplifies the discussion). We prove that the result of the insertion of the stress-energy tensor (the procedure explained above) is described by the Ward identities in the form (\ref{wardzero}), replacing $\Or_1(z_1)\cdots \Or_n(z_n)$ by $\tou$, the set $\{z_1,\ldots,z_n\}\cup \p C$ by $\Sigma\cup \p C$, and correlation functions by probabilities, for $C=\hC$ or $C$ any simply connected domain. The exact statements are in theorems \ref{theowardplane} and \ref{theowarddom}, with definition \ref{T}. These results and their proofs are in clear analogy with the result of \cite{DRC} in the context of SLE$_{8/3}$, for the zero-dimensional fields. They are in a sense more general, since they hold for any event depending on a set $\Sigma$ (which can be, for instance, a set of separated points or a continuous set). But they do not explicitly include multiple insertions of the stress-energy tensor, or Ward identities with fields transforming in other ways than zero-dimensional fields. The proofs of theorems \ref{theowardplane} and \ref{theowarddom}, however, make it very clear that in general we obtain (\ref{wardgen}), with more complicated objects than CLE events that transform in more complicated ways, like the stress-energy tensor itself (but full proofs necessitate a more subtle analysis).

We also obtain a formula for the one-point function of the stress-energy tensor, theorem \ref{theoex1pf}, which relates it to what we call the {\em relative partition function}. It says that the one-point function of the stress-energy tensor at the point $w$ can be obtained from the logarithm of the relative partition function by applying a global holomorphic derivative at $w$, similarly to (\ref{wardgen}). Equivalently, it is obtained by applying a derivative in the direction $h_{w,\theta}$, similarly to (\ref{wardgen3}). This is a purely CLE result, which has no counterpart in SLE$_{8/3}$. In fact, the formula we obtain for the one-point function is what lead us to define the relative partition function of a domain $C$ with respect to another domain $D$ in the CLE context (see definition \ref{defiZ}). It is defined by
\[
	Z(C|D) = \frc{\Preg(\hC\setminus \cl{C})_{\hC}}{\Preg(\hC\setminus\cl{C})_{\hC\setminus\cl{D}}}
\]
for a domain $D$ (the {\em relative domain}) with $\cl{D}\subset C$, and the one-point function of the stress-energy tensor is
\beq
	\bra T(w)\ket_C = \Delta_w^{\hC_w} \log Z(C|D)
\eeq
for $w\in D$. The function $Z(C|D)$ is invariant under global conformal transformations, hence its global holomorphic derivative exists. The derivative $\Delta_w^{\hC_w}$ through which we evaluate the one-point function is with respect to $\p C\cup \p D$. The expression of $\bra T(w)\ket_C$ in terms of the CLE relative partition function seems ``ambiguous'': the relative partition function depends on the relative domain $D$. However, we will show that the derivative of the log of this function with respect to $\p C\cup \p D$ is {\em independent} of $D$ (as long as $w\in D$).

The r\^ole of the domain $D$ in this definition will be explained in the CFT context in section \ref{discuss}. The relation between this formula for the one-point function $\bra T(w)\ket_C$, and the standard CFT formula relating $\bra T(w)\ket_C$ to the variation of the free energy with respect to a metric change, will also be explained there. In a nutshell, the relative partition function $Z(C|D)$ is a particular ratio of ordinary partition functions, that has the property of being invariant under global conformal transformations. Under a metric change that is singular at $w$, essentially what is described by $\Delta_w^{\hC_w}$, the boundary parts of the transformations of the various partition functions in $Z(C|D)$ cancel out, and we are left only with the singular part. This is what puts the stress-energy tensor at $w$.

Finally, we prove that our stress-energy tensor indeed transforms like the CFT stress-energy tensor (\ref{transTCFT}), theorem \ref{theotransfo}, for some central charge. The techniques used there are entirely different from those used in the context of SLE$_{8/3}$ \cite{DRC}. In particular, we obtain a Schwarzian derivative term with a generically non-zero central charge; this occurs through the use of lemma \ref{lemsch}, a general simple result about conformal transformations.

Combining the stress-energy tensor transformation properties and the one-point function formula gives a nice, non-trivial formula for certain ratios of CLE probabilities (or more precisely, for their global holomorphic derivatives). Indeed, since the stress-energy tensor has zero one-point function on the unit disk (a consequence of the fact that it's a second Fourier transform, and that conformal transformations that preserve the disk have zero Schwartzian derivative), we immediately find $\bra T(g(w))\ket_{g(\uD)} = -(c/12)\, \{g,w\}/(\p g(w))^2$. Hence,
\beq\label{formulaPP}
	\Delta_{z}^{\hC_{z}} \log
		\lim_{\varep\to0} \frc{P(\ev(\hC\setminus \cl{C},\varep,u))_{\hC}}{P(\ev(\hC\setminus\cl{C},\varep,u))_{\hC\setminus\cl{D}}}
		= \frc{c}{12} \{s,z\}
\eeq
for any conformal transformation $s: C\to \uD$. Here, we used the formula $\{g,w\} = -\{s,z\} (\p g(w))^2$ where $z=g(w)$ and $s=g^{-1}$. In equation (\ref{formulaPP}), the choice of the function $u$ is arbitrary (see (\ref{eqZ})). With the anti-holomorphic part (assuming that the central charge is real), this gives rise to the conformal derivative formula
\beq
	\nabla_h \log
	\lim_{\varep\to0} \frc{P(\ev(\hC\setminus \cl{C},\varep,u))_{\hC}}{P(\ev(\hC\setminus\cl{C},\varep,u))_{\hC\setminus\cl{D}}}
	= \frc{c}{12} \lt( \oint_{z\in\vec\p (\hC\setminus C)^+} \dd z\, h(z) \{s,z\} +
		\oint_{z\in\vec\p (\hC\setminus C)^+} \bd \b{z}\, \b{h}(\b{z}) \{\b{s},\b{z}\}\rt)
\eeq
for any $h$ holomorphic on (the closed set) $\hC\setminus D$ (with again $\cl{D}\subset C$) except perhaps for a pole of order at most 2 at $\infty$.

Inverting these considerations, the central charge can certainly be written in terms of a derivative of the relative partition function. The value for the central charge that naturally comes out of our calculation is a particular case of such considerations, and is as follows. Consider the inverse of the relative partition function, $1/Z(\hC\setminus \cl{E}|\hC\setminus\cl{D})$, as a function of $\p E\cup \p D$ again. Since its global holomorphic derivative exists, and since it has a bounded partition (in the sense of \cite{diff}), we may consider its charge (\ref{cancharge}) (again in the sense of \cite{diff}). Consider now $E$ to be some standard elliptical domain (see formula (\ref{ellipse}) for the elliptical domain). The central charge $c$ is simply the charge of the logarithm of the inverse relative partition function at $\p E \cup \p D$:
\[
	c = \Gamma\log Z(\hC\setminus\cl{E}|\hC\setminus\cl{D})^{-1}
\]
for any simply connected domain $D$ such that $\cl{D}$ excludes $\infty$ and $\cl{E}\subset D$.

The obtention of the central term in the transformation property of the stress-energy tensor is the most important accomplishment of this paper. As we discuss in section \ref{discuss}, the appearance of a non-zero central charge is indicative of the infinitely many loops around almost every point, present with some ``fixed'' density at all scales. In terms of an underlying statistical model, these loops affect each other in a chain on ``scale space'', and the central charge is the density that emerges from the microscopic interaction when looking at macroscopic scales.

As we mentioned, in the present paper, the only assumptions that we make are those that have to do with differentiability, expressed in assumption \ref{assdiff}. In the first part of this work \cite{I}, Lipschitz continuity was shown for particular events, theorem I.3.6; this can be seen as a small step towards a part of our differentiability assumption.

\sect{Renormalised probabilities}

\subsection{Choice of partners} \label{ssectchoice}

In order to construct the stress-energy tensor, we need to choose, for any given simply connected domain $A$ and $\varep>0$, a fixed event $\ev(A,\varep,u)$; that is, a fixed function $u:\p A \to \hC$. Our choice is guided by the fact that we will require that the stress-energy tensor transforms ``normally'' under {\em global} conformal transformations. The parameter $\varep$ plays the r\^ole of a ``cut-off'', in the language of quantum field theory, and the choice of a fixed $u$ is a choice of a cut-off procedure. Essentially, we choose our cut-off scheme in such a way that global conformal invariance is preserved.

For $\ev(A,\varep,u)$, for any given $A$ and given $\varep$ small enough, we define a unique partner $B$ to $A$, with $\p B = (\id + \varep u) (\p A)$ and $\b{B}$ included inside $A$. In the case where $A = \uD$, the unit disk, we choose $B = (1-\varep) \uD$; that is, $B$ is the disk with radius $1-\varep$. We will denote by $u_\uD$ the function $u$ that reproduces this: $u_\uD(z) = -z$ for $z\in\p\uD$. Hence the event $\ev(\uD,\varep,u_\uD)$ is that no loop transversally cuts the annulus of width $\varep$ with outer boundary $\p\uD$.

Let us denote by $\dom$ the set of all simply connected domains in $\hC$ whose boundaries exclude the point $\infty$. For simplicity, we will also ask that the boundaries be ``smooth enough'': that for $A\in\dom$, any conformal transformation $g:\uD\to A$ is conformal on $\cl\uD$. Let us denote by $\gl$ the set of global conformal transformations. They act on $z\in\hC$ by $\frc{az+b}{cz+d}$ with $ad-bc=1$ where $a,b,c,d$ are in $\C$. Let us denote by $\kl$ the set of transformations, with $SU(1,1)$ group structure, which act on $z\in\hC$ by $\frc{az+\b{b}}{bz+\b{a}}$ with $a\b{a}-b\b{b} = 1$. This is the subgroup of $\gl$ that preserves $\uD$. For any given $A\in\dom$, let us consider the set $[A]_\gl = \{G(A) | G\in\gl\}$. This produces a fibration of $\dom$: if $A\in[A']_\gl$ and $A\in[A'']_\gl$ then $[A']_\gl = [A'']_\gl$, and any element $A$ is in a fiber: $A\in [A]_\gl$. Let us choose a section of this fibration $\se\subset\dom$ such that $\uD\in\se$. That is, $\cup_{A'\in\se} [A']_\gl = \dom$ and $[A]_\gl\cap [A']_\gl=\emptyset$ if $A,A'\in\se$ with $A\neq A'$.

For any $A\in\se$, we fix a conformal map $g_A:\uD\to A$, with in particular $g_\uD = \id$, and define the partner of $A$ as $B=g_A((1-\varep)\uD)$. This is certainly not a unique choice, since $g_A\circ K: \uD\to A$ for any $K\in\kl$, and in general $K((1-\varep)\uD) \neq (1-\varep)\uD$. For any $A'\in [A]_\gl$ with $A\in\se$, we fix a conformal map $G_{A',A}\in\gl$ such that $A'=G_{A',A}(A)$, with in particular $G_{A,A} = \id$, and define $g_{A'} = G_{A',A}\circ g_A$, as well as the partner of $A'$ as $g_{A'}((1-\varep)\uD)$. This is in general also not a unique choice, because $A$ may have a symmetry group: there may be a group $\sym(A)$ of transformations in $\gl$ such that $K(A) = A$ for $K\in\sym(A)$. Two different choices of $G_{A',A}$ are related by such a transformation. For instance, $\sym(\uD) = \kl$, and in general $\sym(A)$ is, as a group, a subgroup of $\kl$. With these choices, we have fixed a map $g_A$ for any $A\in\dom$ such that $A=g_A(\uD)$ and we have chosen the partner of $A$ as $B=g_A((1-\varep)\uD)$.

An important property of these choices is that if $A''=G(A')$ for some global conformal transformation $G\in\gl$, then also their partners are related by a global conformal transformation.
\begin{lemma}\label{lemuglob}
If two simply connected domains $A'$ and $A''$ are related by a global conformal transformation, $A'' = G(A'),\,G\in\gl$, then their partners $B'$ and $B''$ also are, $B'' = \t{G}(B')$ where $\t{G}\in\gl$ such that $A'' = \t{G}(A')$.
\end{lemma}
\proof By construction, we have $g_{A'} = G_{A',A} \circ g_{A}$ and $g_{A''} = G_{A'',A} \circ g_{A}$ for some $A\in\se$, so that $g_{A''} = G_{A'',A} \circ G^{-1}_{A',A}\circ g_{A'} = \t{G}\circ g_{A'}$ where $\t{G}=G_{A'',A} \circ G^{-1}_{A',A}\in\gl$ is such that $A''=\t{G}(A')$. \eproof

The function $u$ on $\p A$ that reproduces our choice of partner will be denoted $u_A$. We have $z+\varep u_A(z) = g_A((1-\varep)g_A^{-1}(z))$, so that $u_A$ in general depends on $\varep$, but uniformly tends to its limit as $\varep\to0$:
\beq
	u_A(z) \to -g_A^{-1}(z) (\p g_A\circ g_A^{-1})(z).
\eeq
By construction, we have
\beq\label{evAD}
    \ev(A,\varep,u_A)_{g_A(B)} = g_A(\ev(\uD,\varep,u_\uD)_B)
\eeq
for any $B$ where $g_A$ is conformal. For $g$ a transformation conformal on $A$, we have $g\circ g_A = g_{g(A)} \circ k$ for some $k\in\kl$. Hence, we have
\beq\label{evA}
    g(\ev(A,\varep,u_A)_B) = \ev(g(A),\varep,g\cdot u_A)_{g(B)}
\eeq
where $A\subseteq B$ and $g$ is conformal on $B$, and where $g\cdot u_A$ is defined through
\beq
	z+\varep (g\cdot u_A)(z) = (g_{g(A)}\circ k)((1-\varep)g_{g(A)}^{-1}(z)).
\eeq
This implies that $g\cdot u_A$ stabilises to
\beq
	-g_{g(A)}^{-1}(z) (\p k\circ g_{g(A)}^{-1})(z) (\p g_{g(A)}\circ k\circ g_{g(A)}^{-1})(z)
\eeq
as $\varep\to0$. Hence, in general $g\cdot u_A \neq u_{g(A)}$; equality occurs if and only if $k={\rm id}$. For global conformal transformations, however, we have a slightly stronger statement, by lemma \ref{lemuglob}: for $G\in\gl$, we have
\beq\label{uglob}
	u_{G(A)} = (G\circ \t{G})\cdot u_A
\eeq
for some global conformal transformation $\t{G}\in\sym(A)$

Our choice of partners for simply connected domains makes it clear that it is impossible to identify $g(\ev(A,\varep,u_A)_B)$ with $\ev(g(A),\varep,u_{g(A)})_{g(B)}$: the function $u$ is in general affected in a different way. This is at the basis of the possibility for a non-zero central charge: the regularisation scheme that we use, characterised by $\varep$, breaks conformal invariance, and the remnant of this breaking survives in the limit $\varep\to0$ to provide the central charge. The necessity of the regulator $\varep$ is a consequence of the presence of infinitely many small loops, hence these are the objects that are seen as responsible for a non-zero central charge, in agreement with the physical intuition. As we mentioned, our choice of regularisation also correctly guarantees that global conformal transformations are not broken; this will be clear below. 

\subsection{Definition of renormalised probabilities}

We are now ready to define {\em renormalised probabilities}, which should be understood as functions similar to probabilities where an event asking that no loop intersect a domain boundary is inserted in conjunction with other events. Naturally, since such an event has exactly zero probability,  many basic properties of probabilities are not expected to be satisfied by renormalised probabilities; for instance, they are not expected to be smaller than or equal to 1. However, this will be a very useful concept, giving rise to an exact {\em restriction} property, instead of the conformal restriction property of CLE, and to a non-trivial {\em conformal covariance}, instead of the conformal invariance property of CLE.

In order to define the renormalised probability, we need the existence of a certain limit.
\begin{propo} \label{proporen}
Consider $\tou$ an event, $A$ a simply connected domain and $C$ a simply connected domain or $C=\hC$. With $\cl{A}\subset C$ and $\supp(\tou)\subset C\setminus \cl{A}$, we have
\beq \label{limren}
	\lim_{\varep\to0} \frc{P(\tou,\ev(A,\varep,u_A))_C}{P(\ev(\uD,\varep,u_\uD))_{2\uD}} \quad \exists.
\eeq
\end{propo}
\proof By theorem I.5.1 (in the case where $C=\hC$) or theorem I.5.2 (in the case where $C$ is a simply connected domain) we have that
\[
    \lim_{\varep\to0} \frc{P(\tou,\ev(A,\varep,u_A))_C}{P(\ev(A,\varep,u_A))_C} \quad\exists
\]
(and it is equal to $P(\tou)_{C\setminus \b{A}}$ by theorem I.5.1 or definition I.5.1). Let us consider $C'\subset C$ small enough (but with $\cl{A}\subseteq C'$) so that $g_A^{-1}$ is conformal on $C'$. Then, by theorem I.5.5,
\[
    \lim_{\varep\to0} \frc{P(\ev(A,\varep,u_A))_C}{P(\ev(A,\varep,u_A))_{C'}} \quad \exists,
\]
and $P(\ev(A,\varep,u_A))_{C'} = P(\ev(\uD,\varep,u_{\uD}))_{g_{A}^{-1}(C')}$ by (\ref{evAD}). Finally, we have that
\[
    \lim_{\varep\to0} \frc{P(\ev(\uD,\varep,u_{\uD}))_{g_{A}^{-1}(C')}}{P(\ev(\uD,\varep,u_\uD))_{2\uD}} \quad\exists
\]
for $C'$ small enough, by theorem I.5.5 again. Multiplying all that, we get (\ref{limren}).
\eproof

From this, we define:
\begin{defi}\label{preg}
The renormalised probability of an event $\tou$ in conjunction with the exclusion of the simply connected domain $A\subset C$, with $\supp(\tou)\subset C\setminus\cl{A}$, and $C$ a simply connected domain or $C=\hC$, is
\beq\label{defreg}
    \Preg(\tou;A)_C =  {\cal N}\lim_{\varep\to0} \frc{P(\tou,\ev(A,\varep,u_A))_C}{P(\ev(\uD,\varep,u_\uD))_{2\uD}}
\eeq
where ${\cal N}>0$ is some number that will be fixed below. If $\tou$ is the trivial event, we will denote the renormalised probability by $\Preg(A)_C$.
\end{defi}
In definition \ref{preg}, it should be remarked that the choice of the denominator is arbitrary to a large extent, as is clear from the presence of the arbitrary finite, strictly positive normalisation constant ${\cal N}$. The unique r\^ole of the denominator is to make the limit exist, thanks to proposition \ref{proporen}.

The steps in the proof of proposition \ref{proporen} gave the renormalised probability as a product of various ratios. The first ratio is $P(\tou)_{C\setminus\cl{A}}$, and the second is $\lim_{\varep\to0} P(\ev(A,\varep,u_A))_C/ P(\ev(A,\varep,u_A))_{C'}$. The other ratios do not depend on $C$. Let us now consider the case where $C$ is a simply connected domain ($\neq \hC$), and look at the limit $\lim_{\lambda\to0} \Preg(\tou;A)_{\lambda_{z',z} C}$ for $z\in C$ and $z'\not\in C$. Recall the notation introduced in \cite{I} for the generalised scale transformation,
\beq\label{genscale}
	\lambda_{z',z}(x) = \frc{(1-\lambda) zz' - (z'-\lambda z) x}{z-\lambda z' - (1-\lambda)x}.
\eeq
For $\lambda$ decreasing, this represents a flow from $z$ to $z'$, which are two fixed point. Re-writing the renormalised probability as a product as above, on the first factor, the limit $\lambda\to0$ exists by theorem I.5.4, and gives $P(\tou)_{\hC\setminus \cl{A}}$. It can also be shown that the limit $\lambda\to0$ on the second factor also exists and gives $\lim_{\varep\to0} P(\ev(A,\varep,u_A))_\hC/ P(\ev(A,\varep,u_A))_{C'}$. The proof of the latter statement simply goes along the lines of the proof of theorem I.5.4 (see \cite{I}), using theorems I.5.5 and I.3.11 instead of I.5.2 and I.3.10, respectively. Putting these factors together, we find
\beq\label{limpreg}
	\lim_{\lambda\to0} \Preg(\tou;A)_{\lambda_{z',z} C} = \Preg(\tou;A)_\hC.
\eeq

Another useful formula for renormalised probabilities is a direct consequence of theorem I.5.5. For $A\subset B\subset C$ and $\p A$, $\p B$, $\p C$ not intersecting each other, this theorem tells us in particular that
\beq
	\lim_{\varep\to 0} \frc{P(\ev(A,\varep,u))_B }{P(\ev(A,\varep,u))_{C}} =
	\lim_{\varep\to 0} \frc{P(\ev(C\setminus\cl{B},\varep,u'))_{C\setminus \cl{A}} }{P(\ev(C\setminus\cl{B},\varep,u'))_C}.
\eeq
With $C=\hC$ and choosing $u=u_A$ and $u' = u_{\hC\setminus \cl{B}}$, we obtain
\beq\label{ratiopreg}
	\frc{\Preg(A)_B}{\Preg(A)_\hC} = \frc{\Preg(\hC\setminus \cl{B})_{\hC\setminus \cl{A}}}{\Preg(\hC\setminus\cl{B})_\hC}.
\eeq

\subsection{Properties of renormalised probabilities}

The first theorem tells us that we have an exact restriction property for renormalised probabilities.
\begin{theorem}\label{theorestreg}
For $C$ a simply connected domain or $C=\hC$, $A$ a simply connected domain and $\tou$ an event supported on $C\setminus\cl{A}$, we have
\beq\label{eqrestreg}
    \frc{\Preg(\tou;A)_C}{\Preg(A)_C} = P(\tou)_{C\setminus \b{A}}
\eeq
\end{theorem}
\proof In the case where $C$ is a simply connected domain, this is an immediate consequence of definitions I.5.1 and \ref{preg}. Indeed, we have, from definition I.5.1,
\beq
	\lim_{\varep\to0} \frc{P(\tou,\ev(A,\varep,u_A))_C}{P(\ev(\uD,\varep,u_\uD))_{2\uD}}
	\frc{P(\ev(\uD,\varep,u_\uD))_{2\uD}}{P(\ev(A,\varep,u_A))_C} = P(\tou)_{C\setminus \b{A}}
\eeq
but the limit exists on both ratios, giving the left-hand side of (\ref{eqrestreg}) by definition \ref{preg}. In the case where $C=\hC$, this follows from theorem I.5.1 in a similar way. \eproof

Second, we have the following important theorem of transformation of renormalised probabilities, which is at the basis of the transformation property of the stress-energy tensor:
\begin{theorem}\label{theotransreg}
For $g$ a transformation conformal on $C$, $g:C\to C'$, with both $C\subset \hC$ and $C'\subset\hC$ simply connected domains, or both $C=\hC$ and $C'=\hC$, for $A$ a simply connected domain with $\cl{A}\subset C$, and for $\tou$ supported on $C\setminus \cl{A}$, we have
\beq
    \Preg(g \tou_C;g(A))_{g(C)} = f(g,A)\Preg(\tou;A)_{C} \label{fctf}
\eeq
where $f(g,A)$ may depend on $g$ and $A$ only.
\end{theorem}
\proof We have, using (\ref{evA}),
\beqa
    \Preg(\tou;A)_{C} &=& \lim_{\varep\to0} \frc{P(\tou,\ev(A,\varep,u_{A}))_{C}}{P(\ev(\uD,\varep,u_\uD))_{2\uD}} \n
        &=& \lim_{\varep\to0} \frc{P(g \tou_C,\ev(g(A),\varep,g\cdot u_{A}))_{g(C)}}{P(g \tou_C,\ev(g(A),\varep,u_{g(A)}))_{g(C)}}
            \frc{P(g \tou_C,\ev(g(A),\varep,u_{g(A)}))_{g(C)}}{P(\ev(\uD,\varep,u_\uD))_{2\uD}}. \no
\eeqa
The second factor on the right-hand side has the finite limit $\Preg(g\tou;g(A))_{g(C)}$ as $\varep\to0$ by definition \ref{preg}, and the product of the two factors also has a finite limit, hence the first also must have a finite limit. We need to prove that this limit is independent of $\tou$ and $C$. First, consider the trivial event for $\tou$. Then we have that
\[
    \lim_{\varep\to0} \frc{P(\ev(g(A),\varep,g\cdot u_{A}))_{g(C)}}{P(\ev(g(A),\varep,u_{g(A)}))_{g(C)}}
\]
exists. In the case where $C\neq\hC$, we can write
\beqa
	&& \lim_{\varep\to0} \frc{P(\ev(g(A),\varep,g\cdot u_{A}))_{g(C)}}{P(\ev(g(A),\varep,u_{g(A)}))_{g(C)}} =\n && \quad =
	\lim_{\varep\to0} \lt[\frc{P(\ev(g(A),\varep,g\cdot u_{A}))_{g(C)}}{P(\ev(g(A),\varep,g\cdot u_{A}))_{\hC}}
	\frc{P(\ev(g(A),\varep,u_{g(A)}))_{\hC}}{P(\ev(g(A),\varep,u_{g(A)}))_{g(C)}}
	\frc{P(\ev(g(A),\varep,g\cdot u_{A}))_\hC}{P(\ev(g(A),\varep,u_{g(A)}))_\hC}\rt] \n && \quad =
	\lim_{\varep\to0}\frc{P(\ev(g(A),\varep,g\cdot u_{A}))_\hC}{P(\ev(g(A),\varep,u_{g(A)}))_\hC}.
\eeqa
In the last step, we have used theorem I.5.5, saying that the limit on the first two factors of the second line exist individually on each factor, and is independent of $g\cdot u_A$ and $u_{g(A)}$, respectively. Since the individual results of the limit are reciprocal to one another, they cancel each other. The result is the same for any $C$. Then, for non-trivial $\tou$, we have
\beqa
	&& \lim_{\varep\to0}
	\frc{P(g \tou_C,\ev(g(A),\varep,g\cdot u_{A}))_{g(C)}}{P(g\tou_C ,\ev(g(A),\varep,u_{g(A)}))_{g(C)}} = \n && \quad =
	\lim_{\varep\to0} \frc{P(g \tou_C|\ev(g(A),\varep,g\cdot u_{A}))_{g(C)}}{P(g\tou_C|\ev(g(A),\varep,u_{g(A)}))_{g(C)}}
	\frc{P(\ev(g(A),\varep,g\cdot u_{A}))_{g(C)}}{P(\ev(g(A),\varep,u_{g(A)}))_{g(C)}} \n && \quad =
	\lim_{\varep\to0}\frc{P(\ev(g(A),\varep,g\cdot u_{A}))_{g(C)}}{P(\ev(g(A),\varep,u_{g(A)}))_{g(C)}} \label{derivation}
\eeqa
where in the last step we used theorem I.5.2 in the case where $C$ is a simply connected domain, or theorem I.5.1 in the case where $C=\hC$. The result is independent of $\tou$ and $C$, which completes the proof.\eproof

Finally, from this and from the choice of partners defining the function $u_A$, described in subsection \ref{ssectchoice}, we can now easily prove global conformal invariance of renormalised probabilities, a property that is crucial in the construction of the stress-energy tensor.
\begin{theorem}\label{invglob}
Renormalised probabilities are invariant under global conformal transformations: for $A$ a simply connected domain, and for $G\subset\gl$, we have
\beq\label{eqinvglob}
    f(G,A) = 1.
\eeq
\end{theorem}
\proof First, note that by (\ref{uglob}),
\beqa
    \Preg(G(A))_{\hC} &=& \lim_{\varep\to0} \frc{P(\ev(G(A),\varep,u_{G(A)}))_{\hC}}{P(\ev(\uD,\varep,u_\uD))_{2\uD}} \n
        &=& \lim_{\varep\to0}\frc{P(G\cdot \t{G} \cdot \ev(A,\varep,u_{A}))_{\hC}}{P(\ev(\uD,\varep,u_\uD))_{2\uD}} \n
        &=& \Preg(A)_\hC \no
\eeqa
for any $G\in\gl$, where $\t{G}\in\sym(A)\subset \gl$. From (\ref{fctf}), we have
\beq
    f(g,A) = \frc{\Preg(g(A))_{g(C)}}{\Preg(A)_{C}}~.
\eeq
Choosing $g=G\in\gl$, we can take $C = \hC$ and we find (\ref{eqinvglob}).
\eproof

\sect{The stress-energy tensor and the conformal Ward identities}

As explained in subsection \ref{ssectCFT}, the stress-energy tensor can be understood, in the realm of CFT, as (the scaling limit of) a certain random variable whose product with other random variables averages to a function with certain analytic property. These averages of products of random variables are correlation functions. Here we find it more convenient to discuss probabilities instead of averages of CLE random variables. Probabilities of a conjunction of events can be seen as averages of the product of their characteristic functions, hence we expect to reproduce correlation functions by considering conjunctions of events. But correlation functions have a multi-linear structure. This naturally translates into linear combinations of probabilities, and of renormalised probabilities. This is why we simply define below the insertion of the stress-energy tensor into a probability with event $\tou$ by a particular linear combination of renormalised probabilities where an additional event is considered in conjunction with $\tou$. We will call this a {\em pseudo-probability}. We will put such constructions in a more general context through the definition of {\em objects} and their correlation functions in section \ref{sectuniv}.

The definition of the stress-energy tensor below is based mainly on the fact that the pseudo-probability representing the insertion of a stress-energy tensor at a point satisfies the correct conformal Ward identities of CFT. This definition, as well as the derivation of the conformal Ward identities from it, parallels very closely what was done in \cite{DRC} in the context of SLE$_{8/3}$, with the notable exception of the equation for the one-point function of the stress-energy tensor and the ensuing definition of the relative partition function. The present derivation uses the renormalised probabilities just introduced instead of ordinary probability. The necessity of using renormalised probabilities comes from the fact that they satisfy a strict conformal restriction, theorem \ref{theorestreg}, contrary to ordinary CLE probabilities. In the construction of \cite{DRC}, this was an essential ingredient, holding for ordinary probabilities in SLE$_{8/3}$. As a consequence, however, strict conformal invariance is lost, and replaced by conformal covariance, expressed in theorem \ref{theotransreg}. This will be at the source of the Schwarzian derivative term in the transformation properties deduced in the next section.

\subsection{CLE definition of the stress-energy tensor and of the relative partition function}

First, consider a simply connected domain, which we will denote by $E(w,\ep,\theta)$, that includes the point $w$ and whose boundary is an ellipse, described by the set of points
\beq\label{ellipse}
	\p E(w,\ep,\theta) = \lt\{w + \ep e^{i\theta} \lt(\frc{b}4 e^{i\alpha} - \frc1{4b} e^{-i\alpha}\rt),\;\alpha\in[0,2\pi)\rt\}~.
\eeq
Here $b>1$ is some parameter, fixed throughout. We use this special domain in order to fix the number ${\cal N}$ of definition $\ref{preg}$, by choosing it in accordance to the following proposition:
\begin{propo}\label{norm}
The number ${\cal N}$ in definition \ref{preg} can be chosen such that for any $w\in \hC$, $w\neq\infty$, any $\ep>0$ and any $\theta\in[0,2\pi]$,
\beq
	\Preg(E(w,\ep,\theta))_\hC = 1.
\eeq
\end{propo}
\proof Simply note that $E(w,\ep,\theta) = e^{i\theta} \ep E(0,1,0)+w$, and use theorem \ref{invglob}:
\beq
	\Preg(E(w,\ep,\theta))_\hC = \Preg(E(0,1,0))_{\hC}.
\eeq
This is finite and independent of $\theta$, $\ep$ and $w$. \eproof

From this choice, the number ${\cal N}$ would vary if we were to change the parameter $b$, but this does not influence any of the considerations below.

The (holomorphic) stress-energy tensor is essentially the ``second Fourier coefficient of the event'' that loops in the elliptical region are separated from the rest in the limit where the ellipse is very small. We will denote the pseudo-probability of $\tou$ in conjunction with the stress-energy tensor at the point $w$ by $P_1(\tou;w)_C$. We define this as follows:
\begin{defi}\label{T}
With $C$ a simply connected domain or $C=\hC$, with $w\in C$ and $w\neq\infty$, and with $\tou$ an event supported in $C$ away from $w$, the pseudo-probability of $\tou$ with the (holomorphic) stress-energy tensor at the point $w$ is
\beq
    P_1(\tou;w)_C := -\lim_{\ep\to0} \frc{8}{\pi \ep^2} \int_0^{2\pi} d\theta e^{-2i\theta} \Preg(\tou;E(w,\ep,\theta))_C.
\eeq
\end{defi}
The index $1$ is introduced because this represents the insertion of only one stress-energy tensor (we hope to study multiple insertions in future works). This definition looks slightly different than that of \cite{DRC} used in the context of SLE$_{8/3}$. However, when definition \ref{T} is specialised to the SLE context, where the renormalised probability is an ordinary probability, it is the same as that of \cite{DRC}. Indeed, the event in \cite{DRC} was that the curve {\em intersects} the ellipse, and the negative sign was absent. But the probability that the curves intersects is 1 minus the probability that the curve does not intersect, and the second Fourier component of 1 is zero. In the CLE context, however, only definition \ref{T} makes sense.

The considerations, below, of the one-point function $P_1(w)_C$ of the stress-energy tensor, and in the next section of the central charge, lead us to the concept of {\em relative partition function} of a domain $C$ with respect to another domain $D$:
\begin{defi} \label{defiZ}
The relative partition function of a simply connected domain $C$ with respect to another simply connected domain $D$ satisfying $\cl{D}\subset C$ is defined by
\beq\label{eqZ}
	Z(C|D) := \frc{\Preg(\hC\setminus \cl{C})_{\hC}}{\Preg(\hC\setminus\cl{C})_{\hC\setminus\cl{D}}} = \frc{\Preg(D)_{\hC}}{\Preg(D)_C}
	= \lim_{\varep\to0} \frc{P(\ev(D,\varep,u))_\hC}{P(\ev(D,\varep,u))_C}.
\eeq
\end{defi}
The first equality in (\ref{eqZ}) is a simple consequence of (\ref{ratiopreg}). The second equality is a consequence of definition \ref{preg}, and the fact that it holds for any $u$ is due to theorem I.5.5. The results that involve the relative partition function will be shown not to depend on $D$. We will discuss this concept further, and give it a CFT interpretation, in section \ref{discuss}.

\subsection{Assumption of differentiability}

In the following we show that definition \ref{T} makes sense (that is, the limit exists), and lead to the conformal Ward identities, as long as the event $\tou$ is ``differentiable''. In order to be more precise, we will consider an event $\tou=\tou(\Sigma)$ that can be associated to a set $\Sigma$, in such a way that its support includes $\Sigma$, and that a transformation $g$ conformal on a domain $B$ including the support gives $g(\tou(\Sigma)_B) = \tou(g(\Sigma))_{g(B)}$. We will require conformal differentiability \cite{diff} (reviewed in subsection \ref{ssectdiff}) of probabilities on $\hC$ as function of $\Sigma$, and of probabilities on domains $C$ as function of $\Sigma \cup \p C$. For simplicity, we will implicitly consider
\[
	\tou = \tou(\Sigma),
\]
without explicitly writing $\Sigma$; also, for a transformation $g$ as above, we will write
\[
	\tou(g(\Sigma)) = g\cdot \tou.
\]

The holomorphic and antiholomorphic $A$-derivatives of conformal differentiability depend on a parameter $a\in\hC\setminus\cl{A}$; we will choose this parameter to be $\infty$ if $\infty\in\hC\setminus\cl{A}$, and any number in $\hC\setminus\cl{A}$ otherwise. Its value does not affect any of the results. In particular, in the cases where the function being differentiated is invariant under global conformal transformation, this choice corresponds to the unique global holomorphic/antiholomorphic $A$-derivatives, which have properties that will be of use, as recalled in subsection \ref{ssectdiff}.

The functions that we differentiate in the theorems below are the probability functions $P(\tou)_C$ as well as the renormalised probabilities $\Preg(E)_C$, for $C$ a simply connected domain or $C=\hC$, and for $E$ a simply connected domain with $\cl{E}\subset C$. They will be seen as functions of $\Sigma\cup \p C$ and $\p E\cup \p C$ respectively; these are the set with respect to which we will differentiate. In the case $C=\hC$, we simply take $\p C = \emptyset$. We will not write explicitly the sets with respect to which we differentiate when there is no ambiguity possible. By conformal invariance of CLE probabilities, and global conformal invariance of renormalised probabilities (theorem \ref{invglob}), $P(\tou)_C$ and $\Preg(E)_C$ are invariant under global conformal transformations. Hence, we will be able to use the special properties of the global holomorphic derivatives.

We make the following assumption:
\begin{assump} \label{assdiff}{\ }
\begin{itemize}
\item The probability $P(\tou)_C$, for $C$ a simply connected domain or $C=\hC$, is $A$-differentiable as a function of $\Sigma\cup \p C$ for any $A$ that contains $\supp(\tou)\cup \p C$. The probability $P(\tou)_{C\setminus \cl{E}}$, for $E$ a simply connected domain with $\cl{E}\subset C$ and $C$ as before, is $A$-differentiable as a function of $\Sigma \cup \p C$ for any $A$ that contains $\supp(\tou) \cup \p C$.
\item Theorem I.5.4 also holds for the derivatives $\Delta_{z\,|\,\tou,\p C}^{A}P(\tou)_{C\setminus \cl{E}}$ and its anti-holomorphic counterpart, in place of $P(\tou)_{C\setminus \cl{E}}$ and with $E$ being scaled down to a point, for any $A$ for which we have $A$-differentiability.
\item The renormalised probability $\Preg(B)_C$, for $C$ a simply connected domain or $C=\hC$ and $B$ a simply connected domain with $\cl{B}\subset C$, is $A$-differentiable as a function of $\p B\cup \p C$ for any $A$ that contains $\p B\cup \p C$. The renormalised probability $\Preg(B)_C$ is also $A$-differentiable as a function of $\p B$ for any $A$ that contains $\p B$.
\item Equation (\ref{limpreg}) also holds for the derivatives $\Delta_{z\,|\,\p B}^{A}\Preg(B)_{C}$ and its anti-holomorphic counterpart, in place of $\Preg(B)_{C}$ and with $C$ being scaled up to $\hC$, for any $A$ for which we have $A$-differentiability.
\end{itemize}
\end{assump}

Note that in \cite{I}, we proved Lipschitz continuity for the events $\ev(A,\varep,u)$ (in the sense of definition I.3.7), theorem I.3.6. With $\tou = \ev(A,\varep,u)$, this is very near to the first part of the first point in assumption \ref{assdiff}. We believe that Lipschitz continuity implies conformal differentiability ``almost everywhere,'' but a further study would be useful. Moreover, proving the other parts of assumption \ref{assdiff} would require more analysis.

\subsection{Conformal Ward identities}

Using differentiability, we will show both that definition \ref{T} makes sense, and that it gives rise to the conformal Ward identities. We proceed in three steps.

First, we consider the case where $C=\hC$.
\begin{theorem}\label{theowardplane}
The limit in definition \ref{T} exists for $C=\hC$, and satisfies the conformal Ward identities on the plane:
\beq
	P_1(\tou;w)_\hC = \Delta_{w}^{\hC_w} P(\tou)_\hC
		\label{wardplane}
\eeq
where $\hC_w = \hC \setminus \cl{N(w)}$ and $N(w)$ is a neighbourhood of $w$ not intersecting $\supp(\tou)$.
\end{theorem}
\proof
Thanks to proposition \ref{norm} and theorem \ref{theorestreg}, we have
\beq\label{pregC}
	\Preg(\tou;E(w,\ep,\theta))_\hC = P(\tou)_{\hC\setminus\cl{E(w,\ep,\theta)}}.
\eeq
Consider the conformal transformation
\beq\label{gwet}
    g_{w,\ep,\theta}(z) = z + \frc{\ep^2 e^{2i\theta}}{16(w-z)}.
\eeq
It is a simple matter to see that $g_{w,\ep,\theta}(\hC\setminus (w+(b\ep/4)\cl\uD)) = \hC\setminus \cl{E(w,\ep,\theta)}$. Hence, we have
\beqa
	\lefteqn{P(\tou)_{\hC\setminus (w+(b\ep/4)\cl\uD)}} && \n
		&=& P(g_{w,\ep,\theta}\cdot \tou)_{\hC\setminus \cl{E(w,\ep,\theta)}} \n &=&
		P(\tou)_{\hC\setminus\cl{E(w,\ep,\theta)}} + \frc{\ep^2}{16} \lt(
		\int_{z\in\vec\p \hC_w} \frc{\dd z\, e^{2i\theta}}{w-z} \Delta_{z\,|\,\tou}^{\hC_w} +
		\int_{z\in\vec\p \hC_w} \frc{\bd\b{z} \,e^{-2i\theta}}{\b{w}-\b{z}} \b\Delta_{\b{z}\,|\,\tou}^{\hC_w}\rt)
			P(\tou)_{\hC\setminus\cl{E(w,\ep,\theta)}} + o(\ep^2) \n &=&
		\Preg(\tou;E(w,\ep,\theta))_\hC + \frc{\ep^2}{16}  \lt(
		\int_{z\in\vec\p \hC_w} \frc{\dd z\, e^{2i\theta}}{w-z} \Delta_{z\,|\,\tou}^{\hC_w} +
		\int_{z\in\vec\p \hC_w} \frc{\bd\b{z} \,e^{-2i\theta}}{\b{w}-\b{z}} \b\Delta_{\b{z}\,|\,\tou}^{\hC_w}\rt)
			P(\tou)_{\hC} + o(\ep^2) \no
\eeqa
where in the first step we used conformal invariance of CLE probabilities, in the second step, we used the first point of assumption \ref{assdiff} and the theory of conformal differentiability, and in the last step, we used the second point. Upon the integration $\int_0^{2\pi} d\theta e^{-2i\theta}$, this gives 
\[
	P_1(\tou;w)_\hC = \int_{z\in\vec\p \hC_w} \frc{\dd z\, }{w-z} \Delta_{z}^{\hC_w} P(\tou)_{\hC}
\]
and in particular the fact that the limit in definition \ref{T} exists in the case $C=\hC$. Equation (\ref{wardplane}) is a consequence of global conformal invariance of $P(\tou)_\hC$: the global holomorphic derivative is holomorphic on $N(w)$, so that the integral can be evaluated. Recall that the integral is counter-clockwise around $\hC_w$, hence it is clockwise around $N(w)$.
\eproof

Second, we infer from the case $C=\hC$ that definition \ref{T} makes sense in the case where $C$ is a simply connected domain, with $\tou$ the trivial event. In order to do so, we ``construct'' the domain $C$ by introducing the event $\ev(C,\varep,u)$ and using theorem I.5.5. Then, a derivation similar to that of theorem \ref{theowardplane} gives well-definiteness of definition \ref{T}. In addition, it provides a formula for the one-point function of the stress-energy tensor. The result is cast into a suggestive form by using our definition \ref{defiZ} of the relative partition function.
\begin{theorem}\label{theoex1pf}
The limit in definition \ref{T} exists for $C$ a simply connected domain and $\tou$ the trivial event, and is equal to
\beq
	P_1(w)_C = \Delta_{w}^{\hC_w} \log Z(C|D)
	\label{p1rest}
\eeq
where $D$ is a simply connected domain such that $w\in D$ and $\cl{D}\subset C$, and $\hC_w = \hC \setminus \cl{N(w)}$ where $N(w)$ is a neighbourhood of $w$ not intersecting $\p D$. The result is independent of the domain $D$.
\end{theorem}
\proof We have from proposition \ref{norm} and equation (\ref{ratiopreg}) (a direct consequence of theorem I.5.5),
\beqa
	\Preg(E(w,\ep,\theta))_C &=& \frc{\Preg(E(w,\ep,\theta))_C}{\Preg(E(w,\ep,\theta))_\hC} \n
	&=& \frc{\Preg(\hC\setminus\cl{C})_{\hC\setminus\cl{E(w,\ep,\theta)}}}{\Preg(\hC\setminus\cl{C})_\hC} \no
\eeqa
From the transformation equation for renormalised probabilities, theorem \ref{theotransreg}, and following the lines of the proof of theorem \ref{theowardplane} with in particular (\ref{gwet}), we find
\beqa
	\lefteqn{f(g_{w,\ep,\theta},\hC\setminus \cl{C}) \Preg(\hC\setminus\cl{C})_{\hC\setminus (w+(b\ep/4)\cl\uD)}} && \n
	&=& \Preg(g_{w,\ep,\theta}(\hC\setminus\cl{C}))_{\hC\setminus \cl{E(w,\ep,\theta)}} \n
	&=& \Preg(\hC\setminus\cl{C})_{\hC\setminus \cl{E(w,\ep,\theta)}} + \n && \quad +
		\frc{\ep^2}{16} \lt(
		\int_{z\in\vec\p \hC_w} \frc{\dd z\, e^{2i\theta}}{w-z} \Delta_{z\,|\,\p C}^{\hC_w} +
		\int_{z\in\vec\p \hC_w} \frc{\bd\b{z} \,e^{-2i\theta}}{\b{w}-\b{z}} \b\Delta_{\b{z}\,|\,\p C}^{\hC_w}\rt)
			\Preg(\hC\setminus\cl{C})_{\hC\setminus\cl{E(w,\ep,\theta)}} + o(\ep^2) \n
	&=& \Preg(\hC\setminus\cl{C})_\hC \Preg(E(w,\ep,\theta))_C + \n && \quad +
		\frc{\ep^2}{16} \lt(
		\int_{z\in\vec\p \hC_w} \frc{\dd z\, e^{2i\theta}}{w-z} \Delta_{z}^{\hC_w} +
		\int_{z\in\vec\p \hC_w} \frc{\bd\b{z} \,e^{-2i\theta}}{\b{w}-\b{z}} \b\Delta_{\b{z}}^{\hC_w}\rt)
			\Preg(\hC\setminus\cl{C})_{\hC} + o(\ep^2). \no
\eeqa
Here, we used the third and fourth point of assumption \ref{assdiff}. Likewise, taking (\ref{fctf}) as a definition of $f$, we can write
\beqa
	\lefteqn{f(g_{w,\ep,\theta},\hC\setminus{\cl{C}})} && \n &=&
		\frc{\Preg(g_{w,\ep,\theta}(\hC\setminus{\cl{C}}))_{g_{w,\ep,\theta}(\hC\setminus{\cl{D}})}}{
			\Preg(\hC\setminus{\cl{C}})_{\hC\setminus{\cl{D}}}} \n
	&=& 1 +
		\frc{\ep^2}{16} \lt(
		\int_{z\in\vec\p \hC_w} \frc{\dd z\, e^{2i\theta}}{w-z} \Delta_{z}^{\hC_w} +
		\int_{z\in\vec\p \hC_w} \frc{\bd\b{z} \,e^{-2i\theta}}{\b{w}-\b{z}} \b\Delta_{\b{z}}^{\hC_w}\rt)
			\log\Preg(\hC\setminus\cl{C})_{\hC\setminus\cl{D}} + o(\ep^2) \no
\eeqa
for any simply connected domain $D$ such that $w\in D$ and $\cl{D}\subset C$. We used the chain rule (\ref{chainrule}) in order to write the derivative term with the logarithmic function. Applying a Fourier transform in $\theta$ and using theorem I.5.4, we find
\[
	-\lim_{\ep\to0} \frc{8}{\pi \ep^2} \int_0^{2\pi} d\theta\, e^{-2i\theta} \, \Preg(E(w,\ep,\theta))_C
		= \Delta_w^{\hC_w} \log \frc{\Preg(\hC\setminus \cl{C})_{\hC}}{\Preg(\hC\setminus\cl{C})_{\hC\setminus\cl{D}}}
\]
so that by definition \ref{defiZ} we obtain (\ref{p1rest}).
\eproof

In (\ref{p1rest}), we can take $\p D$ as near as we want to $\p C$, as long as they do not intersect. Then, by the differentiability assumption and the general theory of conformal derivatives \cite{diff}, $P_1(w)_C$ is holomorphic on $C$. Hence, we can write
\beq\label{p1rest4}
	P_1(w)_C
	= \int_{z\in \vec\p\lt(\hC\setminus \cl{C}\rt)^+} \frc{\dd z}{w-z}
			\Delta_{z}^{\hC_w} \log Z(C|C^-)
\eeq
where the superscript $^+$ indicates that the contour is outside $\hC\setminus \cl{C}$, but infinitesimally close to it, and $\cl{C^-}\subset C$ with $\p C^-$ infinitesimally close to $\p C$.

Finally, we reproduce the derivation of theorem \ref{theowardplane} and use theorem \ref{theoex1pf} in order to obtain the conformal Ward identities in general.
\begin{theorem}\label{theowarddom}
The limit in definition \ref{T} exists for $C$ a simply connected domain, and satisfies the conformal Ward identities on this domain:
\beq
	P_1(\tou;w)_C = P_1(w)_C P(\tou)_C + \Delta_w^{\hC_w} P(\tou)_C
	\label{warddom}
\eeq
where $\hC_w = \hC\setminus \cl{N(w)}$ with $N(w)$ a neighbourhood of $w$ not intersecting $\supp(\tou)\cup \p C$.
\end{theorem}
\proof Consider again the conformal transformation (\ref{gwet}). This time, we see that $g_{w,\ep,\theta}(C\setminus (w+(b\ep/4)\cl\uD)) = g_{w,\ep,\theta}^\sharp(C)\setminus \cl{E(w,\ep,\theta)}$, where $g_{w,\ep,\theta}^\sharp$ is conformal on $C$ and is such that $g_{w,\ep,\theta}^\sharp(\p C) = g_{w,\ep,\theta}(\p C)$. We have:
\beqa
	\lefteqn{P(\tou)_{C\setminus (w+(b\ep/4)\cl\uD)}} && \n &=&
		P(g_{w,\ep,\theta}\cdot \tou)_{g_{w,\ep,\theta}^\sharp(C)\setminus \cl{E(w,\ep,\theta)}} \n &=&
		P(\tou)_{C\setminus\cl{E(w,\ep,\theta)}}  + \frc{\ep^2}{16} \int_{z\in\vec\p \hC_w} \lt(
		\frc{\dd z\, e^{2i\theta}}{w-z} \Delta_{z\,|\,\tou,\p C}^{\hC_w} +
		\frc{\bd\b{z} \,e^{-2i\theta}}{\b{w}-\b{z}} \b\Delta_{\b{z}\,|\,\tou,\p C}^{\hC_w}\rt)
			P(\tou)_{C\setminus\cl{E(w,\ep,\theta)}} + o(\ep^2) \n &=&
		\frc{\Preg(\tou;E(w,\ep,\theta))_{C}}{\Preg(E(w,\ep,\theta))_C}
		+ \frc{\ep^2}{16} \int_{z\in\vec\p \hC_w} \lt(
		\frc{\dd z\, e^{2i\theta}}{w-z} \Delta_{z\,|\,\tou,\p C}^{\hC_w} +
		\frc{\bd\b{z} \,e^{-2i\theta}}{\b{w}-\b{z}} \b\Delta_{\b{z}\,|\,\tou,\p C}^{\hC_w}\rt)
			P(\tou)_{C} + o(\ep^2). \no
\eeqa
In the first step, we used conformal invariance of CLE probabilities on annular domains, theorem I.5.3. In the second step, we used the first point of assumption \ref{assdiff}. In the last step, we used the second point of assumption \ref{assdiff} as well as theorem \ref{theorestreg}. From (\ref{limpreg}) and global conformal invariance theorem \ref{invglob}, we see that $\lim_{\ep\to0} \Preg(E(w,\ep,\theta))_C = \Preg(E(w,1,\theta))_\hC$ which is $1$ by proposition \ref{norm}. Hence we find
\beqa
	&& \Preg(E(w,\ep,\theta))_C P(\tou)_{C\setminus (w+(b\ep/4)\b\uD)} = \n && \qquad \Preg(\tou;E(w,\ep,\theta))_{C}
	+ \frc{\ep^2}{16} \int_{z\in\vec\p \hC_w} \lt(
		\frc{\dd z\, e^{2i\theta}}{w-z} \Delta_{z}^{\hC_w} +
		\frc{\bd\b{z} \,e^{-2i\theta}}{\b{w}-\b{z}} \b\Delta_{\b{z}}^{\hC_w}\rt)
			P(\tou)_{C} + o(\ep^2). \no
\eeqa
Upon the integration $\int_0^{2\pi} d\theta e^{-2i\theta}$, using theorems \ref{theoex1pf} and I.5.4, we obtain
\[
	P_1(\tou;w)_C = P_1(w)_C P(\tou)_C + \int_{z\in \vec\p \hC_w}\frc{\dd z\, }{w-z} \Delta_{z}^{\hC_w} P(\tou)_{C}
\]
and then (\ref{warddom}).
\eproof

It is possible to write the insertion of the stress-energy tensor purely as a global holomorphic derivative. We have to use, for this purpose, not the probabilities $P(\tou)_C$, but the probabilities multiplied by the factor $Z(C|D)$. In the CFT language, this essentially corresponds to using somewhat ``un-normalised'' correlation functions, where the relative partition function has not been normalised away. We get:
\beq\label{wardZ}
	Z(C|D) P_1(\tou;w)_C = \Delta_w^{\hC_w} \lt[Z(C|D) P(\tou)_C\rt].
\eeq
It is interesting to remark that from this, it should be possible to understand the Virasoro algebra of the modes of the stress-energy tensor through multiple global holomorphic derivatives of the quantity $Z(C|D) P(\tou)_C$.

In our derivation, we have used conformal invariance of CLE probability functions, and we have assumed that the event $\tou$ was characterised by a set $\Sigma$ that transforms into $g(\Sigma)$ when a conformal transformation is applied on $\tou$. In the CFT language, this essentially corresponds to the case where the fields are dimensionless, equation (\ref{wardzero}); except that in the CLE situation, we do not need $\Sigma$ to be a finite set of points. However, it is clear from the proofs that cases where the transformation properties are more complicated can be done in an entirely equivalent way. That is, suppose our starting point is not a probability function $P(\tou)_C$ but another object, like a renormalised probability itself, in which we want to insert a stress-energy tensor. If this object transforms non-trivially under transformations conformal on $C$, the result of the insertion of the stress-energy tensor is still expressed through the global holomorphic derivative, (\ref{wardplane}) and (\ref{warddom}), but now applied to the object seen as a function of conformal transformations, in the same way as in equation (\ref{wardgen}).

Note finally that by rotational symmetry, we have
\beq\label{oneptuD}
	P_1(0)_\uD=0.
\eeq
That is, the one-point function of the stress energy tensor is zero when it is at the center of the unit disk (or of any disk). The results of the next section imply that it is also zero at any point inside a disk (as usual in CFT).

\sect{Transformation of the stress-energy tensor}

Contrary to some of the derivations of the previous section, the derivation of the transformation properties of the stress-energy tensor in the present context constitutes a major departure from the derivation presented in \cite{DRC} in the case of the stress-energy tensor in SLE$_{8/3}$. This is because from the principles used in \cite{DRC}, there was essentially no way of obtaining a transformation that included a non-zero central charge. The presence of a non-zero central charge comes from the subtleties of CLE as compared to SLE$_{8/3}$, and obtaining it was one of the main reasons for investigating the construction in the context of CLE.

The transformation properties of the stress-energy tensor follow from two effects. One is that a conformal transformation of the elliptical domain, if we look at the second Fourier component in the limit where the ellipse is very small, is equivalent to a translation, a rotation and a scaling transformation, up to an additional Schwarzian derivative term. The derivation of this effect is based on a re-derivation of the conformal Ward identities, as in theorem \ref{theowarddom}, for an elliptical domain affected by a conformal transformation, and on a proposition about the change of normalisation that occurs when the ellipse is transformed. The second effect is that of the ``anomalous'' transformation properties of the renormalised probabilities, theorem \ref{theotransreg}. The factor involved in the transformation property, (\ref{fctf}), can be evaluated and gives rise to another Schwarzian derivative contribution. Then, in total, the stress-energy tensor transforms by getting a factor of the derivative-squared of the conformal transformation, plus a Schwarzian derivative; this is the usual transformation property in conformal field theory.

Note that for global conformal transformations, the Schwarzian derivative is zero, so that the stress-energy tensor transforms like a field of dimension $(2,0)$. This can in fact be directly deduced from the expressions (\ref{wardplane}), (\ref{p1rest}) and (\ref{warddom}), as from the general theory of conformal derivatives \cite{diff}, the global holomorphic derivative transforms in this way for global conformal transformations, equation (\ref{cantrans}). Further, from (\ref{warddom}), the ``connected part'' $P_1(\tou;w)_C - P_1(w)_C P(\tou)_C$ transforms like a field of dimension $(2,0)$ under {\em any} transformation conformal on $C$, thanks to the property (\ref{cantrans2}) of the global holomorphic derivative. However, we will not need to deduce these transformation properties in this way: the property (\ref{cantrans2}) necessitates a slightly stronger differentiability assumption (that is nevertheless expected to hold), and in any case, considering the connected part does not provide the central charge. Our method deals directly with the CLE definition of $P_1(\tou;w)_C$.

The involvement of the Schwarzian derivative, in conformal field theory, is usually understood through the unique finite transformation equation associated to infinitesimal generators forming the Virasoro algebra. For instance, the Schwarzian derivative term in the finite transformation equation is proportional to the central charge of the Virasoro algebra. These infinitesimal generators are the modes (coefficients of the doubly-infinite power series expansion) of the stress-energy tensor, and their algebra can be derived from the conformal Ward identities, when many insertions of the stress-energy tensor are considered. In the present paper, we do not study this algebra, or multiple insertions of the stress-energy tensor (we hope to come back to these subjects in future works). The Schwarzian derivative is obtained independently from the Virasoro algebra structure underlying the multiple-insertion conformal Ward identities. The basis for its appearance in our calculations is the following simple result in the theory of conformal transformations:
\begin{lemma}\label{lemsch}
Given a transformation $g$ conformal in a neighbourhood of $w\neq\infty$, there is a unique global conformal transformation $G$ such that
\beq\label{condsch}
	(G\circ g)(z)  = z + \frc{a}6 (z-w)^3 + O((z-w)^4)
\eeq
for some coefficient $a$, and this coefficient is uniquely determined by $g$ and $w$ to be
\beq\label{sch}
	a = \{g,w\} := \lt\{ \ba{ll}
		\frc{\p^3 g(w)}{\p g(w)} - \frc32 \lt(\frc{\p^2g(w)}{\p g(w)}\rt)^2 & (g(w)\neq\infty) \z
		-3\lim_{z\to w} \lt(\frc{2 \p g(z)}{(z-w)g(z)} + \frc{\p^2 g(z)}{g(z)} \rt)& (g(w) = \infty).
	\ea\rt.
\eeq
In the case $g(w)\neq\infty$, this is the usual Schwarzian derivative of $g$ at $w$. In the other case, this should be understood as a definition of the Schwarzian derivative.
\end{lemma}
\proof Let us first consider the case $g(w)\neq\infty$. Then, since $g$ is conformal around $w$, we have $\p g(w)\neq0$, and we can write
\[
	g(z) = g(w) + \p g(w) (z-w) + \frc{\p^2 g(w)}2 (z-w)^2 + \frc{\p^3 g(w)}{3!} (z-w)^3 + O((z-w)^4).
\]
It is convenient to construct $G$ in two steps. First, we may eliminate the constant and linear terms through a unique combination of a translation, rotation and scaling transformation:
\[
	g(z) = (G_1\circ h)(z),\quad G_1(z) = g(w) + \p g(w)(z-w),\quad h(z) = z + h_2(z-w)^2 + h_3(z-w)^3 + O((z-w)^4)
\]
with
\[
	h_2 = \frc{\p^2 g(w)}{2\p g(w)},\quad h_3 = \frc{\p^3g(w)}{3!\p g(w)}.
\]
Second, the only global conformal transformations that do not involve local translations, rotations and scaling around $w$ are of the form
\[
	w + \frc{z-w}{1+\eta (z-w)} = z - \eta(z-w)^2 + \eta^2(z-w)^3 + O((z-w)^4)
\]
for some $\eta\in\C$. In $(G_2\circ h)(z)$ for $G_2$ of that form, the requirement that the power $(z-w)^2$ disappears uniquely fixes $\eta = h_2$, so that we find
\[
	(G_2\circ G_1^{-1} \circ g)(z) = z + (h_3-h_2^2)(z-w)^3 + O((z-w)^4)
\]
which reproduces (\ref{sch}) in the case $g(w)\neq\infty$.

Then, let us consider $g(w)=\infty$. Since $g$ is conformal around $w$, it must have an expansion
\[
	g(z) = \frc{A}{z-w} + B + C(z-w) + O((z-w)^2)
\]
with $A\neq0$. Let us apply a global conformal transformation $z\mapsto 1/z$. We obtain
\beq\label{1gz}
	\frc{1}{g(z)} = \frc{1}A (z-w) - \frc{B}{A^2} (z-w)^2 + \lt(\frc{B^2}{A^3}-\frc{C}{A^2}\rt) (z-w)^3 + O((z-w)^4).
\eeq
We can then use the result just established, with $\t{g}(z) = 1/g(z)$ in place of $g(z)$. This immediately gives the case $g(w)=\infty$ of equation (\ref{sch}).
\eproof

\subsection{Contribution from the transformation of a small elliptical domain}

The Schwarzian derivative appears when we consider the transformation of a small elliptical domain in the renormalised probability on $\hC$:
\begin{propo}\label{propotrren}
For $g$ a transformation conformal in a neighbourhood of $w\neq\infty$, we have
\beq\label{eqtrren}
	\Preg(g(E(w,\ep,\theta)))_\hC = 1 - \frc{\ep^2}{192}\lt( e^{2i\theta} \{g,w\} c_1 +
		e^{-2i\theta} \{\b{g},\b{w}\} \b{c}_1\rt) + o(\ep^2)
\eeq
where
\beq\label{c1}
	c_1 = -\chrg \log\Preg(E(0,1,0))_\hC, \quad
	\b{c}_1 = -\b\chrg \log \Preg(E(0,1,0))_\hC.
\eeq
\end{propo}
\proof Using the global conformal transformation $G$ of theorem \ref{lemsch} and global conformal invariance, theorem \ref{invglob}, we have
\beqa
	\Preg(g(E(w,\ep,\theta)))_\hC &=& \Preg((G\circ g)(E(w,\ep,\theta)))_\hC \n
		&=& \Preg(\ep^{-1} e^{-i\theta} ((G\circ g)(\ep e^{i\theta} E(0,1,0)+w)-w))_\hC. \no
\eeqa
From lemma \ref{lemsch}, it is easy to see that
\beq\label{eqexpl}
	\ep^{-1} e^{-i\theta} ((G\circ g)(\ep e^{i\theta} z+w)-w) = z + \ep^2 h_{w,\ep,\theta}(z)
\eeq
where $h_{w,\ep,\theta}(z)$ converges uniformly to $\{g,w\} e^{2i\theta} z^3/6$ as $\ep\to0$ for any $z$ in compact subsets of the finite complex plane. Hence we find
\[
	\Preg(g(E(w,\ep,\theta)))_\hC = \Preg((\id + \ep^2 h_{w,\ep,\theta})(E(0,1,0)))_\hC \n
\]
which gives (\ref{eqtrren}) with
\beqa
	c_1 &=& -32\int_{z\in \vec\p \hC_\infty} \dd z\,z^3 \Delta_{z}^{\hC_\infty} \Preg(E(0,1,0))_\hC, \\
	\b{c}_1 &=& -32\int_{z\in \vec\p \hC_\infty} \bd\b{z}\,\b{z}^3 \b\Delta_{\b{z}}^{\hC_\infty} \Preg(E(0,1,0))_\hC
\eeqa
by differentiability (the third point of assumption \ref{assdiff}) and by the normalisation given in proposition \ref{norm}. Here $\hC_\infty = \hC\setminus \cl{N(\infty)}$ where $N(\infty)$ is a neighbourhood of $\infty$ not intersecting the elliptical domain $E(0,1,0)$. We can perform the integral by evaluating the pole at $z=\infty$, using (\ref{cancharge}). Dividing by $\Preg(E(0,1,0))_\hC=1$ and re-writing the result through the chain rule (\ref{chainrule}) for convenience, this gives (\ref{c1}).
\eproof

In order to obtain the transformation properties of the stress-energy tensor, it is natural to study the second Fourier component of renormalised probabilities, as occurs in the definition \ref{T}, but where the domain excluded is an elliptical domain that is affected by a conformal transformation $g$. We show that it is related to the same object with the ellipse kept untransformed, times the factor $(\p g(w))^2$, up to an additional Schwarzian derivative term. The factor $(\p g(w))^2$ comes from the fact that the elliptical domain is affected by the local translation, rotation and scaling of the conformal transformation, and the Schwarzian derivative factor comes from the change of normalisation described by proposition \ref{propotrren}.
\begin{propo}\label{propotrell}
For $C$ a simply connected domain or $C=\hC$, $w\in C$ with $w\neq\infty$, and $\tou$ an event supported in $C$ away from $w$, and for $g$ a transformation conformal on a domain containing $w$ with $g(w)\neq \infty$, we have
\beq\label{trell}
	-\lim_{\ep\to0} \frc{8}{\pi\ep^{2}} \int_0^{2\pi} d\theta e^{-2i\theta} \Preg(\tou;g(E(w,\ep,\theta)))_C
	= (\p g(w))^2 P_1(\tou;g(w))_C + \frc{c_1}{12} \{g,w\} P(\tou)_C.
\eeq
\end{propo}
\proof First, we can write $g = G\circ h$ where $G(z) = g(w) + \p g(w) (z-w)$, and $h=z + O((z-w)^2)$. Then, it is sufficient to prove that
\beqa
&&	-\lim_{\ep\to0} \frc{8}{\pi\ep^{2}} \int_0^{2\pi} d\theta e^{-2i\theta} \Preg(\tou;h(E(w,\ep,\theta)))_C = \n
&&	\qquad = -\lim_{\ep\to0} \frc{8}{\pi\ep^{2}} \int_0^{2\pi} d\theta e^{-2i\theta} \Preg(\tou;E(w,\ep,\theta))_C + \frc{c_1}{12}\{h,w\}P(\tou)_C.
\label{intermtrell}
\eeqa
Indeed, if we have (\ref{intermtrell}), we then find
\beqa
&&	-\lim_{\ep\to0} \frc{8}{\pi\ep^{2}} \int_0^{2\pi} d\theta e^{-2i\theta} \Preg(\tou;g(E(w,\ep,\theta)))_C = \n
&&	\qquad = -\lim_{\ep\to0}\frc{8}{\pi\ep^{2}} \int_0^{2\pi} d\theta e^{-2i\theta} \Preg(G^{-1}\tou;h(E(w,\ep,\theta)))_{G^{-1}(C)} \n
&&	\qquad = -\lim_{\ep\to0}\frc{8}{\pi\ep^{2}} \int_0^{2\pi} d\theta e^{-2i\theta} \Preg(G^{-1}\tou;E(w,\ep,\theta))_{G^{-1}(C)} +
		\frc{c_1}{12}\{h,w\}P(\tou)_C \n
&&	\qquad = -\lim_{\ep\to0} \frc{8}{\pi\ep^{2}}\int_0^{2\pi} d\theta e^{-2i\theta} \Preg(\tou;E(g(w),|\p g(w)|\ep,\theta+{\rm arg}(\p g(w))))_{C}
		+ \frc{c_1}{12}\{h,w\}P(\tou)_C \n
&&	\qquad = - (\p g(w))^2 \lim_{\ep\to0} \frc{8}{\pi\ep^{2}}
	\int_0^{2\pi} d\theta e^{-2i\theta} \Preg(\tou;E(g(w),\ep, \theta))_C
	+ \frc{c_1}{12} \{g,w\}P(\tou)_C,
\eeqa
where in the last step we used the existence of the limit, theorems \ref{theowardplane} or \ref{theowarddom}, and the fact that $\{h,w\} = \{G\circ h,w\}$ for any global conformal transformation $G$, thanks to lemma \ref{lemsch}. This is the desired result.

In order to prove (\ref{intermtrell}), we follow the steps of the proofs of theorems \ref{theowardplane}, \ref{theoex1pf} and \ref{theowarddom}. Hence, we are seeking, in replacement of $g_{w,\ep,\theta}$ (\ref{gwet}), a conformal transformation $\t{g}$ as follows:
\beq\label{seeking}
	\t{g}(\hC\setminus(w+(b\ep/4)\cl\uD)) = \hC \setminus h(\cl{E(w,\ep,\theta)}).
\eeq
Since the boundary of the transformed elliptical domain $h(E(w,\ep,\theta))$ is smooth, $\t{g}$ is in fact conformal on $\hC\setminus(w+(b\ep/4)\uD)$ (that is, it is conformal on a domain containing the closed set $\hC\setminus(w+(b\ep/4)\uD)$). Let us use the variable $v=e^{i\alpha},\,\alpha\in[0,2\pi)$ in order to parametrise the unit circle, and
\beq
	z(v) = w + \ep e^{i\theta} \lt(\frc{b}4 v - \frc1{4bv}\rt)
\eeq
in order to parametrise the ellipse (see (\ref{ellipse})). Clearly, $h$ makes modifications of order $O(\ep^2)$ to this boundary:
\beq\label{corr}
	h(z(v)) = w + \ep e^{i\theta} \lt(\frc{b}4 v - \frc1{4b} v^{-1}\rt) + O(\ep^2)
\eeq
where $O(\ep^2)$ is uniform in $v$. The condition (\ref{seeking}) is equivalent to asking that $\t{g}$ be conformal on $\hC\setminus(w+(b\ep/4)\uD)$, and that
\beq\label{condtg}
	\t{g}(w+(b\ep/4)\t{v}) = h(z(v))
\eeq
where $v\mapsto\t{v}$ is a change of parametrisation of the unit circle.

For $\ep$ small enough, $h(E(w,\ep,\theta))$ does not contain the point $\infty$. Then, we may make the map $\t{g}$ unique up to a rotation about $w$ by asking that it fixes the point $\infty$. We may further completely fix it by requiring that the coefficient of the term in $(z-w)$ in an expansion about $\infty$ is positive:
\beq
	\t{g}(z) = \t{g}_1(z-w) + w + \t{g}_0 + \sum_{m\leq -1} \t{g}_m (z-w)^m,\quad \t{g}_1>0
\eeq
where in general $\t{g}_m$ depend on $w,\ep,\theta$. With this choice, the map $v\mapsto\t{v}$ is also unique. Note that these requirements are satisfied by $g_{w,\ep,\theta}$ (with in particular $\t{g}_1=1,\,\t{g}_0=0$), so that for $h=\id$ we recover $\t{g} = g_{w,\ep,\theta},\,\t{v}=ve^{i\theta}$.

Instead of $\t{g}$, let us consider
\[
	\tau(z) = \frc{\t{g}(\ep z+w)-w}{\ep} = \sum_{m\leq 1} \t{g}_m \ep^{m-1} z^m.
\]
It is the unique conformal transformation on $\hC\setminus (b/4)\uD$ that maps the circle $|z|=b/4$ to a the deformed ellipse $(h(z(v))-w)/\ep$ ($v\in\p\uD$), that preserves the point $\infty$, and whose coefficient of $z$ in an expansion about $\infty$ is positive. As $\ep\to0$, we find that the deformed ellipse becomes the usual ellipse up to terms $O(\ep)$ uniformly in $v$:
\[
	\frc{h(z(v))-w}{\ep} = e^{i\theta} \lt(\frc{b}4 v - \frc1{4b} v^{-1}\rt) + O(\ep).
\]
Hence,
\[
	\tau(z) = \frc{g_{w,\ep,\theta}(\ep z+w)-w}{\ep} + O(\ep) = z - \frc{ e^{2i\theta}}{16z} + O(\ep)
\]
where $O(\ep)$ is uniform for $z$ in any compact subset of $\hC\setminus (b/4) \cl\uD$. With contour integrals, we can isolate the coefficients $\t{g}_m$, and we find that $\t{g}_1 = 1+\delta_1$, $\t{g}_0 =\delta_0$, $\t{g}_{-1} = -\ep^2 e^{2i\theta}/16 + \delta_{-1}$ and $\t{g}_m = \delta_m$ for $m\leq -2$, with $\delta_m = O(\ep^{2-m})$ for $m\leq 1$. It is possible to evaluate $\delta_m$ order by order in $\ep$. We simply have to find a reparametrisation of the unit circle $v\mapsto \t{v}$ such that $(h(z(v))-w)/\ep$ has a Fourier expansion in the first, zeroth and negative powers of $\t{v}$ only. We may well choose $v=\t{v}e^{-i\theta} + O(\ep)$, so that $\tau(z)$ is obtained by making the replacement $(b/4) \t{v} \mapsto z$ in this expansion. Upon further imposing that $\t{g}_1>0$, this guarantees that $\tau(z)$ has the correct analytic properties, and the correct boundary conditions, $\tau((b/4)\t{v} ) = (h(z(v))-w)/\ep$. To first order in $\ep$, this transformation can be shown to have the form
\[
	v = \frc{(1+\alpha \ep)\t{v}e^{-i\theta} + \beta \ep}{\b\beta \ep \t{v}e^{-i\theta} + 1 + \b\alpha \ep}
\]
for some complex numbers $\alpha$ and $\beta$. A calculation indeed shows that we can satisfy all conditions to first order in $\ep$, as long as $\b\beta = (b/4)\p^2 h(w)$ (in order not to have the $z^2$ power in $\tau(z)$) and $\alpha = \b\alpha$ (in order that $\t{g}_1$ be real). This calculation also provides $\delta_1$ and $\delta_0$ to leading order in an expansion in powers of $\ep$ (to first order for $\delta_1$, and to second order for $\delta_0$). In particular, we have $\delta_1 = O(\ep^2)$; that is, it is in fact zero to first order in $\ep$. We will make use of this below.

Let us now write
\[
	\delta(z) = \sum_{m\leq -1} \delta_m (z-w)^m.
\]
We have that $\delta(z)= O(\ep^3)$ uniformly for any $z$ on compact subsets a finite distance away from $w$. With this notation, we can write $\t{g}(z)$ as
\beq\label{tgG}
	\t{g}(z) = G \lt(z + \frc{\ep^2 e^{2i\theta}}{16(w-z)} +
		\frc{\delta(z) - \frc{\delta_1 \ep^2 e^{2i\theta}}{16(w-z)}}{1+\delta_1} \rt)
\eeq
where $G(z) = (1+\delta_1) z +  \delta_0 - \delta_1 w$ is a combination of a translation and a scaling with respect to the point $w$. From the results above, this may be written $G = \id + \ep^2 H_{\ep^2}$ where $H_{\ep^2}$ converges uniformly (let's say to a function $H$) as $\ep\to0$ on any compact subset of $\hC_w$. Similarly, by the previous considerations, the argument of $G$ in (\ref{tgG}) has the form $\id + \ep^2 q_{\ep^2}$ where $q_{\ep^2}$ is a holomorphic function on $\hC_w$ which converges uniformly as $\ep^2\to0$ to the function $z\mapsto e^{2i\theta}/(16(w-z))$ on any compact subset of $\hC_w$.

Then, re-tracing the steps of the proof of theorem \ref{theowardplane}, using conformal invariance of CLE probabilities, differentiability (first point of assumption \ref{assdiff}), conformal restriction (theorem \ref{theorestreg}), and finally the second point of assumption \ref{assdiff}, we have
\beqa
	\lefteqn{P(\tou)_{\hC\setminus (w+(b\ep/4)\cl\uD)}}\n
		 &=& P(\t{g}\cdot \tou)_{\hC\setminus h(\cl{E(w,\ep,\theta)})} \n &=&
		\frc{\Preg(G\tou;h(E(w,\ep,\theta)))_\hC}{\Preg(h(E(w,\ep,\theta)))_\hC} + \frc{\ep^2}{16}  \lt(
		\int_{z\in\vec\p \hC_w} \frc{\dd z\, e^{2i\theta}}{w-z} \Delta_{z}^{\hC_w} +
		\int_{z\in\vec\p\hC_w} \frc{\bd\b{z} \,e^{-2i\theta}}{\b{w}-\b{z}} \b\Delta_{\b{z}}^{\hC_w}\rt)
			P(\tou)_{\hC} + o(\ep^2). \no
\eeqa
In order to deal with the transformation $G$, we first moved it to the domain of definition, then wrote the expansion in terms of conformal derivatives, and moved it back to the event $\tou$ in the zeroth order term. Multiplying through by $\Preg(h(E(w,\ep,\theta)))_\hC$ and applying a Fourier transform, we find
\beqa
	\lefteqn{\frc{8}{\pi \ep^2} \int d\theta\,e^{-2 i\theta}\lt(
	\Preg(h(E(w,\ep,\theta)))_\hC P(G^{-1}\tou)_{\hC\setminus (w+(b\ep/4)\cl\uD)} -\Preg(\tou;h(E(w,\ep,\theta)))_\hC \rt)} &&
	\hspace{15cm} \n
	&=&
	\int_{z\in\vec\p \hC_w} \frc{\dd z\,}{w-z} \Delta_{z}^{\hC_w} P(G^{-1}\tou)_{\hC} + o(1) \n
	&=& \Delta_w^{\hC_w} P(G^{-1}\tou)_\hC
	+ o(1) \label{eqabc}
\eeqa
where we used the fact that the global holomorphic derivative is holomorphic on $N(w)$. Using its transformation property (\ref{cantrans}), the fact that $G=\id + O(\ep^2)$ means that $G$ can just be omitted on the right-hand side of the last equation. The factor $P(G^{-1}\tou)_{\hC\setminus (w+(b\ep/4)\cl\uD)}$ can be partly evaluated to
\beqa
	P(G^{-1}\tou)_{\hC\setminus (w+(b\ep/4)\cl\uD)} &=& P(\tou)_{\hC\setminus (w+(b\ep/4)\cl\uD)} -
		\ep^2 \nabla_{H\,|\,\tou} P(\tou)_{\hC\setminus (w+(b\ep/4)\cl\uD)} + o(\ep^2) \n
	&=& P(\tou)_{\hC\setminus (w+(b\ep/4)\cl\uD)} -
		\ep^2 \nabla_{H\,|\,\tou} P(\tou)_{\hC} + o(\ep^2) \n
	&=& P(\tou)_{\hC\setminus (w+(b\ep/4)\cl\uD)} + o(\ep^2) \no
\eeqa
where in the first two steps we used the first two points of assumption \ref{assdiff}, and in the last step we used global conformal invariance. Then, from (\ref{eqabc}), using proposition \ref{propotrren} and theorems \ref{theowardplane} and I.5.4, we obtain (\ref{intermtrell}) in the case $C=\hC$.

In order to obtain the case where $\tou$ is the trivial event and $C$ is a simply connected domain, we may re-trace the steps of the proof of theorem \ref{theoex1pf}. From proposition \ref{norm}, equation (\ref{ratiopreg}) and proposition \ref{propotrren}, we start with
\beqa
	\lt(1 + \frc{\ep^2}{192} \lt(e^{2i\theta} \{h,w\} c_1 + c.c.\rt)\rt) \Preg(h(E(w,\ep,\theta)))_C + o(\ep^2)
	&=& \frc{\Preg(h(E(w,\ep,\theta)))_C}{\Preg(h(E(w,\ep,\theta)))_\hC} \n
	&=& \frc{\Preg(\hC\setminus\cl{C})_{\hC\setminus h(\cl{E(w,\ep,\theta)})}}{\Preg(\hC\setminus\cl{C})_\hC}. \no
\eeqa
Then, from the transformation equation for renormalised probabilities, theorem \ref{theotransreg}, and following the lines of the proof of theorem \ref{theoex1pf}, we find
\beqa
	\lefteqn{f(\t{g},\hC\setminus \cl{C}) \Preg(\hC\setminus\cl{C})_{\hC\setminus (w+(b\ep/4)\cl\uD)}} && \n
	&=& \Preg(\t{g}(\hC\setminus\cl{C}))_{\hC\setminus h(\cl{E(w,\ep,\theta)})} \n
	&=& \Preg(G(\hC\setminus \cl{C}))_{\hC\setminus h(\cl{E(w,\ep,\theta)})} + \n && \quad +
		\frc{\ep^2}{16} \lt(
		\int_{z\in\vec\p \hC_w} \frc{\dd z\, e^{2i\theta}}{w-z} \Delta_{z}^{\hC_w} +
		\int_{z\in\vec\p \hC_w} \frc{\bd\b{z} \,e^{-2i\theta}}{\b{w}-\b{z}} \b\Delta_{\b{z}}^{\hC_w}\rt)
			\Preg(\hC\setminus\cl{C})_{\hC} + o(\ep^2). \no
\eeqa
We used global conformal invariance of renormalised probabilities, theorem \ref{invglob}, in order to move $G$ around. We also used the third and fourth points of assumption \ref{assdiff}. We then partly evaluate $\Preg(G(\hC\setminus \cl{C}))_{\hC\setminus h(\cl{E(w,\ep,\theta)})}$ as follows:
\beqa
	\lefteqn{\Preg(G(\hC\setminus \cl{C}))_{\hC\setminus h(\cl{E(w,\ep,\theta)})}} && \n &=&
		\Preg(\hC\setminus \cl{C})_{\hC\setminus h(\cl{E(w,\ep,\theta)})} +
		\ep^2 \nabla_{H\,|\,\p C} \Preg(\hC\setminus \cl{C})_{\hC\setminus h(\cl{E(w,\ep,\theta)})} + o(\ep^2) \n
	&=&
		\Preg(\hC\setminus \cl{C})_{\hC\setminus h(\cl{E(w,\ep,\theta)})} +
		\ep^2 \nabla_{H\,|\,\p C} \Preg(\hC\setminus \cl{C})_{\hC} + o(\ep^2) \n
	&=&
		\Preg(\hC\setminus \cl{C})_{\hC\setminus h(\cl{E(w,\ep,\theta)})} +
		o(\ep^2) \no
\eeqa
where in the first two steps we used the third and fourth point of assumption \ref{assdiff}, and in the last step we used theorem \ref{invglob}. Hence,
\beqa
	\lefteqn{f(\t{g},\hC\setminus \cl{C}) \Preg(\hC\setminus\cl{C})_{\hC\setminus (w+(b\ep/4)\cl\uD)}} && \n
	&=& \lt(1 + \frc{\ep^2}{192} \lt(e^{2i\theta} \{h,w\} c_1 + c.c.\rt)\rt) \Preg(h(E(w,\ep,\theta)))_C \Preg(\hC\setminus\cl{C})_\hC
		+ \n && \quad +
		\frc{\ep^2}{16} \lt(
		\int_{z\in\vec\p \hC_w} \frc{\dd z\, e^{2i\theta}}{w-z} \Delta_{z}^{\hC_w} +
		\int_{z\in\vec\p \hC_w} \frc{\bd\b{z} \,e^{-2i\theta}}{\b{w}-\b{z}} \b\Delta_{\b{z}}^{\hC_w}\rt)
			\Preg(\hC\setminus\cl{C})_{\hC} + o(\ep^2). \label{eqcde}
\eeqa
Since $f(G\circ g,A) = f(g,A)$ for any global conformal transformation $G$, the result (\ref{tgG}) as well as differentiability (third and fourth points of assumption \ref{assdiff}) imply that $f(\t{g},\hC\setminus \cl{C}) = f(g_{w,\ep,\theta},\hC\setminus \cl{C}) + o(\ep^2)$. Also, we have
\[
	\lim_{\ep\to0} \Preg(h(E(w,\ep,\theta)))_C =
	\lim_{\ep\to0} \Preg((\ep^{-1}_{w,\infty}\circ  h \circ \ep_{w,\infty})(E(w,1,\theta)))_{\ep^{-1}_{w,\infty} C}
	= \Preg(E(w,1,\theta))_\hC = 1
\]
where we use the generalised scale transformation (\ref{genscale}), theorem \ref{invglob}, equation (\ref{limpreg}) and proposition \ref{norm}. Putting these last two results together, the Fourier transform of equation (\ref{eqcde}) exactly reproduces (\ref{intermtrell}) in the case where $\tou$ is the trivial event and $C$ is a simply connected domain.

Finally, we can do the general case in much the same way as in the case $C=\hC$ above, following the notation and the proof of theorem \ref{theowarddom}. We have
\beqa
	\lefteqn{P(\tou)_{C\setminus (w+(b\ep/4)\cl\uD)}}\n
		 &=& P(\t{g}\cdot \tou)_{\t{g}^\sharp (C)\setminus h(\cl{E(w,\ep,\theta)})} \n &=&
		\frc{\Preg(G\tou;h(E(w,\ep,\theta)))_{G(C)}}{\Preg(h(E(w,\ep,\theta)))_{G(C)}} + \frc{\ep^2}{16}  \lt(
		\int_{z\in\vec\p \hC_w} \frc{\dd z\, e^{2i\theta}}{w-z} \Delta_{z}^{\hC_w} +
		\int_{z\in\vec\p\hC_w} \frc{\bd\b{z} \,e^{-2i\theta}}{\b{w}-\b{z}} \b\Delta_{\b{z}}^{\hC_w}\rt)
			P(\tou)_{C} + o(\ep^2) \no
\eeqa
and then
\beqa
	\lefteqn{\frc{8}{\pi \ep^2} \int d\theta\,e^{-2 i\theta}\lt(
	\Preg(h(E(w,\ep,\theta)))_C P(G^{-1}\tou)_{G^{-1}(C)\setminus (w+(b\ep/4)\cl\uD)} -\Preg(\tou;h(E(w,\ep,\theta)))_C \rt)}
	&& \hspace{15cm} \n
	& = &
	\int_{z\in\vec\p \hC_w} \frc{\dd z\,}{w-z} \Delta_{z}^{\hC_w} P(G^{-1}\tou)_{G^{-1}(C)} + o(1) \n
	&=& \Delta_w^{\hC_w} P(G^{-1}\tou)_{G^{-1}(C)}
	+ o(1) \n &=& \Delta_w^{\hC_w} P(\tou)_{C}
	+ o(1). \no
\eeqa
We partly evaluate $P(G^{-1}\tou)_{G^{-1}(C)\setminus (w+(b\ep/4)\cl\uD)}$ using the first two points of assumption \ref{assdiff} as well as global conformal invariance:
\beqa
	P(G^{-1}\tou)_{G^{-1}(C)\setminus (w+(b\ep/4)\cl\uD)} &=& P(\tou)_{C\setminus (w+(b\ep/4)\cl\uD)} -
		\ep^2 \nabla_{H\,|\,\p C,\tou} P(\tou)_{C\setminus (w+(b\ep/4)\cl\uD)} + o(\ep^2) \n
	&=& P(\tou)_{C\setminus (w+(b\ep/4)\cl\uD)} -
		\ep^2 \nabla_{H\,|\,\p C,\tou} P(\tou)_{C} + o(\ep^2) \n
	&=& P(\tou)_{C\setminus (w+(b\ep/4)\cl\uD)} + o(\ep^2) \no
\eeqa
and (\ref{intermtrell}) follows using theorems \ref{theowarddom} and I.5.4.
\eproof

\subsection{Contribution from the anomalous transformation properties of renormalised probabilities}

The other contribution to the transformation property of the stress-energy tensor comes from that of the renormalised probabilities, theorem \ref{theotransreg}. In order to identify it, we need to study $f(g,A)$ defined by (\ref{fctf}), and in particular $f(g,E(w,\ep,\theta))$.

An insight can be gained into $f(g,A)$ in general by noticing that it is an automorphic factor for the group of conformal transformations:
\beq\label{auto}
	f(h\circ g,A) =
		\frc{\Preg((h\circ g)(A))_{(h\circ g)(C)}}{\Preg(g(A))_{g(C)}} \frc{\Preg(g(A))_{g(C)}}{\Preg(A)_{C}} = f(g,A) f(h,g(A)).
\eeq
Consider $f(g,E(w,\ep,\theta))$. By the symmetries of the elliptical domain, we certainly have
\beq
    f(g,E(w,\ep,\theta)) = \sum_{n\in\Z} f_{2n}(g,w,\ep)e^{2ni\theta}.
\eeq
Also, from (\ref{limpreg}) and proposition \ref{norm}, using the fact that $g$ becomes, locally around $w$, just a combination of a translation, a rotation and a scale transformation and using global conformal invariance, it is possible to show that
\beq
    \lim_{\ep\to0} f(g,E(w,\ep,\theta)) = 1.
\eeq
Through a slightly more precise analysis of the $\theta$-dependence of the leading small-$\ep$ terms of $\Preg(E(w,\ep,\theta))_{C}$, it is possible to argue from the definition of $f(g,A)$ that
\beq
    f_2(g,w,\ep) = \ep^2 f_2(g,w) + o(\ep^2)
\eeq
and that all other Fourier components are of higher order in $\ep$, except for the zeroth component. Hence, we find the infinitesimal version of (\ref{auto}),
\beq\label{defeq}
    f_2(h\circ g,w) = f_2(g,w) + (\p g(w))^2 f_2(h,g(w)).
\eeq
This equation is what is usually obtained in CFT when considering the finite transformation properties of the stress-energy tensor. A solution is the Schwarzian derivative; with additional assumptions, this solution may be made unique (up to normalisation).

This derivation is very natural, but it requires a proof of uniqueness of the solution to (\ref{defeq}). Instead, we will employ a more direct route, deriving the main properties of $f(g,E(w,\ep,\theta))$ through a calculation similar to that of proposition \ref{propotrren}. The Schwarzian derivative naturally comes out from this calculation. We show the following proposition:
\begin{propo}\label{propoexpf}
For $g$ conformal on a neighbourhood of $w\neq\infty$, we have
\beq\label{expf}
	f(g,E(w,\ep,\theta)) = 1 + \frc{\ep^2}{192}\lt( e^{2i\theta} \{g,w\} c_2 + e^{-2i\theta} \{\b{g},\b{w}\} \b{c}_2\rt) + o(\ep^2)
\eeq
where
\beq\label{c2}
	c_2 = \chrg\log \Preg(E(0,1,0))_D, \quad
	\b{c}_2 = \b\chrg\log \Preg(E(0,1,0))_D
\eeq
for any simply connected domain $D$ such that $\cl{D}$ excludes $\infty$ and such that $\cl{E(0,1,0)}\subset D$. The numbers $c_2$ and $\b{c}_2$ are independent of $D$.
\end{propo}
\proof Using (\ref{fctf}) with $\tou$ the trivial event, we have
\beq
	f(g,E(w,\ep,\theta)) = \frc{\Preg(g(E(w,\ep,\theta)))_{g(C)}}{\Preg(E(w,\ep,\theta))_{C}}
\eeq
for any $C$ such that $\cl{E(w,\ep,\theta)}\subset C$, $\infty\not\in \cl{C}$ and such that $g$ is conformal on $C$ (which can be achieved for $\ep$ small enough). Let us choose $C=\ep e^{i\theta} D + w$ for some $D$ such that $\cl{E(0,1,0)}\subset D$ -- this is a valid choice for all $\ep>0$ (small enough so that $g$ is conformal on $C$), since $E(w,\ep,\theta) = \ep e^{i\theta} E(0,1,0) + w$. We may analyse the numerator using lemma \ref{lemsch}. Let us denote by $G$ the global conformal transformation associated to $g$, as in the lemma. Equation (\ref{eqexpl}) along with global conformal invariance immediately implies
\[
	\Preg(g(E(w,\ep,\theta)))_{g(C)} = \Preg((\id + \ep^2 h_{w,\ep,\theta})(E(0,1,0)))_{(\id + \ep^2 h_{w,\ep,\theta})(D)}.
\]
Since the denominator is simply $\Preg(E(w,\ep,\theta))_{C} = \Preg(E(0,1,0))_D$ by global conformal invariance,
we obtain (\ref{expf}) with
\beqa
	c_2 &=& 32 \int_{z\in \vec\p \hC_\infty} \dd z\,z^3 \Delta_z^{\hC_\infty} \log \Preg(E(0,1,0))_D, \n
	\b{c}_2 &=& 32 \int_{z\in \vec\p \hC_\infty} \bd\b{z}\,\b{z}^3 \b\Delta_{\b{z}}^{\hC_\infty} \log \Preg(E(0,1,0))_D \no
\eeqa
by differentiability (the third point of assumption \ref{assdiff}). Here $\hC_\infty = \hC\setminus \cl{N(\infty)}$ where $N(\infty)$ is a neighbourhood of $\infty$ not intersecting $D$. Performing the integral by taking the residue at $\infty$ given by (\ref{cancharge}), we find (\ref{c2}). Finally, since $f(g,E(w,\ep,\theta))$ is independent of $D$ for any $\theta$, a Fourier transform shows that the expressions for $c_2$ and $\b{c}_2$ are also independent of $D$.
\eproof

\subsection{Final transformation equation}

Finally, we may put together propositions \ref{propotrell} and \ref{propoexpf} in order to obtain the final transformation equation for the stress-energy tensor.
\begin{theorem}\label{theotransfo}
For $C$ a simply connected domain or $C=\hC$, $w\in C$ with $w\neq \infty$, $\tou$ an event supported on $C$ away from $w$, and $g$ a transformation conformal on $C$, we have
\beq\label{eqtransfo}
    (\p g(w))^2 P_1(g\cdot \tou;g(w))_{g(C)} + \frc{c}{12} \{g,w\} P(\tou)_{C}  = P_1(\tou;w)_C
\eeq
where
\beq\label{central}
	c = c_1 + c_2 = \chrg \log Z(\hC\setminus\cl{E(0,1,0)}|\hC\setminus\cl{D})^{-1}
\eeq
for any simply connected domain $D$ such that $\cl{D}$ excludes $\infty$ and such that $\cl{E(0,1,0)}\subset D$. The number $c$ is independent of $D$. Here, $\chrg$ is an operator applied on $\log Z(E'|D')$ seen as a function of $\p E'\cup \p D'$.
\end{theorem}
\proof From (\ref{fctf}), we have
\beq
    \int_0^{2\pi} d\theta e^{-2i\theta} f(g,E(w,\ep,\theta)) \Preg(\tou;E(w,\ep,\theta))_{C} = \int_0^{2\pi} d\theta e^{-2i\theta}
    \Preg(g\cdot \tou;g(E(w,\ep,\theta)))_{g(C)}.
\eeq
Using theorem \ref{theorestreg}, theorem I.5.4, equation (\ref{limpreg}) and proposition \ref{norm}, we see that
\[
	\lim_{\ep\to0} \Preg(\tou;E(w,\ep,\theta))_C = P(\tou)_C.
\]
Then, with proposition \ref{propoexpf} on the left-hand side, and proposition \ref{propotrell} on the right-hand side, we find
\[
	c = \chrg\log \frc{\Preg(E(0,1,0))_D}{\Preg(E(0,1,0))_\hC}
\]
and definition \ref{defiZ} gives (\ref{eqtransfo}).
\eproof

The constant $c$ in (\ref{eqtransfo}) is the central charge of the conformal field theory. As mentioned before, its meaning in the Virasoro algebra would be obtained by studying multiple insertions of the stress-energy tensor.

Naturally, combining the transformation property of the stress-energy tensor with the expression for the one-point function (\ref{p1rest}), one could obtain different expressions for the central charge $c$ than that given in (\ref{central}). It is possible, however, to check that expression (\ref{central}) is consistent with the stress-energy tensor one-point function. Consider the domain $\hC\setminus (b/4) \cl{\uD}$, and the transformation $g(z) = z-1/(16 z)$. This transforms the domain into $\hC\setminus \cl{E(0,1,0)}$. Since we must have $P_1(w)_{\hC\setminus (b/4) \cl{\uD}}=0$, the transformation property (\ref{eqtransfo}) gives
\[
	\lt(1+\frc1{16 w^2}\rt)^2 P_1(g(w))_{\hC\setminus \cl{E(0,1,0)}} = -\frc{8c}{(1+16w^2)^2}.
\]
From (\ref{p1rest}), this gives us an expression for $\Delta_w^{\hC_\infty} \log Z(\hC\setminus \cl{E(0,1,0)}|\hC\setminus \cl{D})$, and with simple algebra we find that the large-$w$ expansion is given by $-(c/32) w^{-4} + O(w^{-5})$, in agreement with (\ref{central}).

\sect{Universality and correlation functions}\label{sectuniv}

The transformation property (\ref{eqtransfo}) along with the one-point function (\ref{oneptuD}) on the unit disk allows one to evaluate the one-point function $P_1(w)_C$ for any $w$ and any simply connected domain $C$ using conformal transformations. In particular, these pseudo-probabilities are completely fixed by the central charge $c$ (\ref{central}). From (\ref{p1rest}), $P_1(w)_C$ is expressed purely in terms of a derivative of a ratio of renormalised probabilities that does not involve the elliptical domain $E(w,\ep,\theta)$, neither the normalisation constant ${\cal N}$ involved in definition \ref{preg}. Hence, it must be that $c$ is independent of our particular choice of eccentricity for the ellipse, that is, of the constant $b$ in (\ref{ellipse}):
\begin{corol}
The constant $c$ in (\ref{central}) is universal: it is independent from the eccentricity of the ellipse, i.e.\ of the parameter $b$ introduced in (\ref{ellipse}).
\end{corol}
Then, from the conformal Ward identities, we have a universal definition of the stress-energy tensor: any choice of $b$ gives the same pseudo-probabilities $P(\tou;w)_C$.

In fact, one could perhaps imagine using an object different from the elliptical domain; it is possible that the derivations above could be generalised. Additionally, there are other objects for which the transformation properties are as those of the stress-energy tensor, and one may wonder if they do correspond to different representations of the same stress-energy tensor.

We argue below that any object that transforms like the stress-energy tensor and that is zero on the disk, satisfies the conformal Ward identities. Hence it is a representation of the stress-energy tensor, the same stress-energy tensor if it transforms with the same central charge. In order to make the statement in generality, we need concepts of objects and their correlation functions, inspired by the results about the stress-energy tensor. These should be related to fields and their correlation functions in CFT.
\begin{defi}\label{object}
An object $\Or$ is a two-parameter family of events $\{\tou(t;\ep),\;0<\ep<\ep_0,\,t\in[0,1)\}$ for some $\ep_0$, and a family of functions (or distributions) $s(\ep):[0,1)\to \C,\;t\mapsto s(t;\ep)$. The support of an object $\supp(\Or)$ is a closed set such that for any closed set $B$ that does not intersect $\supp(\Or)$, there exists a $\ep'>0$ such that $B$ does not intersect $\cup_{t\in[0,1)}\supp(\tou(t;\ep))$ for all $0<\ep<\ep'$. Correlation functions of objects are defined by limits of linear combinations of probabilities:
\beq
    \bra \Or_1 \cdots \Or_n\ket_C = \lim_{\ep_1\to0,\ldots,\ep_n\to0} \int_0^1 dt_1\cdots dt_n
    s_1(t_1;\ep_1) \cdots s_n(t_n;\ep_n) P(\tou_1(t_1;\ep_1),\ldots,\tou_n(t_n;\ep_n))_C
\eeq
for $\supp(\Or_i)$ disjoint for different values of $i$ and included inside $C$. A set of objects is a consistent set if and only if all such limits exist (for domains $C\subset\hC$ and for $C=\hC$) and are independent of the order in which they are taken. Likewise, we can form objects of ``second order'', out of families of objects instead of families of events.
\end{defi}
An event is an object (consisting of a ``family'' of the same event, independent of $t$ and $\ep$). Any set of events is then a consistent set of objects. The family $\Or_A = \{\ev(A,\varep,u_A),\;\varep>0,t\in[0,1)\}$ is an object, with $s(t;\varep)=1/P(\ev(\uD,\varep,u_{\uD}))_{2\uD}$, and it forms a consistent set of objects with any set of events. The renormalised probabilities are the corresponding correlation functions. The stress-energy tensor is an object of the ``second order'': it is the family $T(w) = \{\Or_{E(w,\ep,\theta)},\;\ep>0,\,\theta\in[0,2\pi)\}$, with $s(\theta;\ep) = 8 e^{-2i\theta}/(\pi \ep^2)$. The set composed of the stress-energy tensor and any set of events is also a consistent set of objects. The stress-energy tensor at $w$ has support $w$.

\noindent{\bf Statement of universality}{\em\quad
Consider a family composed of events and some object $\Or(w)$. If the object is supported on a point $w$ and defined for any $w\in\C$, is zero on the disk $\uD$, and transforms like the stress-energy tensor $T(w)$ with the same central charge, then it satisfies the conformal Ward identities in that family. }

\noindent{\em Sketch of proof.} Let us denote by $\Or_\tou$ the object associated to some event $\tou$. The proof goes by proving that the correlation functions $\bra \Or(w) \Or_\tou\ket_C$ are equal to $\bra T(w) \Or_\tou \ket_C$ for any $\tou$ supported away from $w$ and for any simply connected domain $C\subset\hC$. Once this is established, the case $C=\hC$ is obtained by the definition of probabilities on $\hC$, definition I.4.2. Let us consider first the case where $\tou$ is the empty event. Then, the equality is just a consequence of the transformation properties, as is discussed above. Otherwise, let us consider a sample of configurations and evaluate the correlation function on each configuration. By the nesting property of CLE, we may take into consideration the object $T(w)$ by evaluating its average in the domain bounded by a loop surrounding $w$ and separating it from $\supp(\tou)$ (see subsection 2.4 of \cite{I}), since there is almost surely such a loop. We may do the same if we have $\Or(w)$ instead of $T(w)$. This evaluation gives the same result in both cases, since it only depends on the central charge and the shape of the domain. Hence, both correlation functions are equal. \eproof

\sect{Discussion}\label{discuss}

\subsection{Regularisation and renormalisation}

An important concept in constructing QFT from a microscopic model is that of regularisation and renormalisation. From the viewpoint of a lattice model, the lattice is seen as a regularisation, and the scaling limit (see the brief discussion in subsection \ref{ssectCFT}) is the renormalisation, leading to well-defined QFT fields and correlation functions. However, in general, many different microscopic models, with many different regularisation-renormalisation procedures, can lead to the same QFT model (this is QFT universality).

The CLE construction of the stress-energy tensor presented here involved essentially two regularisation-renormalisation steps. First, we needed to define renormalised probabilities, probabilities that no loop intersect the boundary of a given domain, this being taken in conjunction with any proper events in the sigma-field. The former event is ill-defined, and in order to obtain a non-zero, meaningful result, we used a prescribed regularisation, a ``fattening'' the boundary of the domain, and a renormalisation, dividing the probability by a factor and taking the limit where the fattening becomes zero. Here, the fattening of the boundary played the role of the lattice spacing, or any other regularisation procedure, where short-distance or large-energy modes are cut off; it cuts off the contributions of small loops. The renormalisation procedure was obtained by simply changing the normalisation of the probability (multiplicative renormalisation), so that the limit where the cut-off (the fattening) is sent to zero is finite and well-defined. Second, from this renormalised probability, we defined the stress-energy tensor by a further regularisation-renormalisation process. We took the second Fourier transform of a renormalised probability with the condition that no loop intersect a small ellipse; the Fourier transform is with respect to the angle the ellipse makes with some fixed direction. Here, the extent of the ellipse is the regularisation parameter. Then, we took the normalised limit of this object as the ellipse becomes very small; this is the renormalisation process.

These two steps led to two contributions to the central charge (the first one is $c_2$, proposition \ref{propoexpf}, the second is $c_1$, proposition \ref{propotrren}). In each case, the contribution can be seen to occur because of the presence of the infinitely many small loops around any point. Indeed, in the first case, it arises due to the anomalous transformation property of renormalised probabilities, involving the coefficient $f(g,A)$, theorem \ref{theotransreg}. This coefficient comes out because a conformal transformation changes the fattening of the boundary of the domain $A$ in a way that is, in general, in disagreement with the prescribed fattening defining the renormalised probability. In other words, the regularisation explicitly breaks conformal invariance, and this breaking subsists in the limit where the regularisation parameter goes to zero. This is a common phenomenon in QFT, where symmetries of the ``classical'' continuum model are broken by quantum fluctuations. Our particular choice of fattening, however, guaranteed that global conformal transformations are not broken. This agrees with the usual wisdom of CFT, according to which classically one has the full infinite-dimensional algebra of infinitesimal conformal transformations, but in the quantum version, local conformal symmetries are broken, only global conformal symmetries subsist (leading to the Virasoro algebra, the central extension of the Witt algebra). The second contribution to the central charge, $c_1$, came from the residual terms in transforming a small elliptical domain in $\hC$: not only the elliptical domain gets translated, rotated and scaled, but there is an additional deformation which cannot be taken away by global conformal invariance. This deformation affects the normalisation of the renormalised probability, which was chosen so that $P(E(0,1,0))_\hC=1$. The necessity of a normalisation is due to the necessity of taking the limit of small fattening, again a consequence of the presence of the small loops.

\subsection{Point splitting and a different representation of the stress-energy tensor}

Our construction of the stress-energy tensor is somewhat similar to the ``point-splitting'' construction in the free boson model (with $c=1$). There, renormalised ``free fields'' are first defined through some QFT regularisation followed by an appropriate renormalisation, then the stress-energy tensor is defined by a product of two (holomorphic derivative of) free fields subtracted by a constant, in the limit where they approach each other (additive renormalisation). In this case, however, the anomalous transformation of the stress-energy tensor (that is, the Schwarzian derivative) comes only from the additive renormalisation, since the free field themselves transform only as dimension-$(1,0)$ fields. Hence, there is only one ``important'' regularisation-renormalisation step. Our construction is different in that in the definition of the stress-energy tensor, we have a multiplicative renormalisation, and there are two ``independent'' contributions to the the anomalous transformation properties, in both regularisation-renormalisation steps. Also, in our two-step construction, the first step does not lead to a proper ``local'' quantity: the renormalised probabilities are associated to boundaries of domains. However, our approach is perhaps the most appropriate for the stress-energy tensor and its descendants, since renormalised probabilities, through their geometric character, can be directly connected to Ward identities associated to space symmetries.

Yet, through the statement of universality in section \ref{sectuniv}, it is possible to construct the stress-energy tensor in a way that is very close to the free-field construction, but valid for any central charge. It involves only one regularisation-renormalisation step, with an additive renormalisation. Indeed, consider the random variable $n(z_1,z_2)$ counting $k$ times the number of loops that surround both points $z_1$ and $z_2$, for some $k>0$. As $z_1\to z_2$, the average of this random variable diverges logarithmically (since a change of scale by a fixed amount increases the number of loops by a fixed amount, in average). This random variable should be identified, intuitively, with a product of free fields in the CFT language, and $|z_1-z_2|$ with the point-splitting regularisation. Hence, let us consider
\[
	\Or(w) = \lim_{|z_1-z_2|\to0} \p_{z_1} \p_{z_2} \lt(n(z_1,z_2) - \frc{c}2 \log|z_1-z_2|\rt)
\]
where the limit is taken with $(z_1+z_2)/2 = w$ fixed (note that this can be seen as an object according to our general definition \ref{object}). With $c$ chosen properly, this limit, when evaluated inside probability functions, is finite. This is the renormalised product of derivatives of free fields, and, with an appropriate choice of $k$, should be identified with the stress-energy tensor. One can see that it is supported at the point $w$. The variable $n(z_1,z_2)$ is not supported on $\{z_1,z_2\}$, because if a loop surrounds the two points, we cannot count the number of loops just by looking inside this loop; the support is in fact $\hC$. However, the variation with respect to $z_1$, for instance, can be obtained just by looking inside a surrounding loop, hence the derivatives are supported on $\{z_1,z_2\}$, and in the limit the support is $w$. Furthermore, with an appropriate choice of $k$, it is possible to make $c$ equal to the central charge (\ref{central}). Then, we can repeat the standard derivation of CFT showing that it transforms like the stress-energy tensor with appropriate central charge\footnote{I would like to thank J. Cardy for sharing with me some time ago a closely related idea for constructing an object with this transformation property.}, using the fact that $g(n(z_1,z_2)) = n(g(z_1),g(z_2))$ for a conformal transformation $g$. Hence, by the statement of universality, it also satisfies the conformal Ward identities.

\subsection{Nonnegativity of the central charge, and the case $\kappa=8/3$}

Here we provide a heuristic argument suggesting that the CLE central charge (as we defined it) should be nonnegative, and in particular should be zero at $\kappa=8/3$.

The main observation is that the loops near the boundary tend to be smaller than those away from it. This is simple to see, on the disk for instance, from conformal invariance: in non-compact directions of the symmetry group, loops in the bulk get nearer to the boundary, and smaller. In a similar spirit, loops near the boundary should also be ``scarcier,'' since no loop can touch the boundary. Let us consider the definition \ref{T} of the stress-energy tensor in CLE. In the renormalised probability involved, we take the width of the ellipse (the boundary of the elliptical domain) to zero in a prescribed manner. But the prescription guarantees that the width is unaffected by rotations, since these are global conformal transformations: at various angles, it is the ``same'' fattened ellipse that we have. Hence we may compare the probability that no loop crosses a certain part of the fattened ellipse at various angles. If this part of the ellipse is near to the domain boundary, the probability that no loops crosses it should be greater. Now let us consider the conformal transformation
\[
	g(z) = \frc{z}{b+z^2/b}
\]
for some $b>1$. It maps the unit disk $\uD$ to a domain that is elongated in the vertical direction. According to the transformation property (\ref{eqtransfo}), and evaluating the Schwarzian derivative, we have simply $P_1(0)_{g(\uD)} = c/2$. Now, if the principal axis of the elliptical domain in the definition \ref{T} is aligned with $g(\uD)$, the renormalised probability should be smaller, as all parts of the elliptical domain are as far as possible from the boundary. On the other hand, if it is not aligned, then the renormalised probability should be greater. Since it is aligned for the angles $\theta=0,\,\pi$ of the elliptical domain and perpendicular for $\theta=\pi/2,\,3\pi/2$, and since we must integrate with the phase $-e^{-2i\theta}$, we find that the integral should be positive. Hence, we find $c>0$.

In the case where $\kappa$ is sent to 8/3, we should obtain a central charge equal to zero. The theory with $\kappa=8/3$ is essentially that of the single self-avoiding loop \cite{W05b}. In this case, there is no problem in defining a probability that the loop does not intersect a domain boundary, so that renormalised probabilities are just ordinary probabilities. These satisfy exact conformal restriction, like renormalised probabilities, but also exact conformal invariance; the factor $f(g,A)$ in (\ref{fctf}) is 1. This means that the contribution $c_2$ of proposition \ref{propoexpf} is zero. Moreover, on $\hC$, the probability that the loop does not intersect a given domain boundary is 1, since the loop is almost surely away from it (there is too much space in $\hC$ for a single loop). Hence, the contribution $c_1$ of proposition \ref{propotrren} is also zero. That is, we indeed find that the central charge as we defined it is zero at $\kappa=8/3$.

\subsection{Renormalised probabilities and partition functions}

In principle, it is not clear {\sl a priori} that the stress-energy tensor that we constructed in the present paper is the {\em correct} one. The three elements that allowed us to identify the stress-energy tensor are the conformal Ward identities, the fact that the Schwarzian derivative is involved in its transformation property, and the fact that is one-point function is zero on the disk. With these three elements, the only remaining parameter that determines all correlation functions involving the stress-energy tensor on simply connected domains is the central charge. Hence, having the correct stress-energy tensor means having the correct central charge. Our construction does not guarantee that the central charge that we defined is the one expected from the CFT central charge of the underlying $O(n)$ model \cite{N82}; or the one expected from the stochastic CLE construction \cite{W05a,ShW07} (both being expected to agree). For instance, perhaps our construction gives a central charge equal to zero, so that we would essentially only have {\em connected} correlation functions of the stress-energy tensor -- a trivial result. Also, in general, if we do not have the correct stress-energy tensor, we can always add a term affecting only its one-point function on simply connected domains, in such a way that the central charge is shifted to the correct one.

In this subsection, we argue that our construction gives the correct stress-energy tensor. Certainly, the universality principle of section \ref{sectuniv} would not apply if a spurious term needed to be added: it was important that our construction correspond to a local CLE object. Furthermore, below we provide strong arguments showing that the expression for the one-point function (\ref{p1rest}), and in particular the relative partition function in definition \ref{defiZ}, are in agreement with general CFT principles. Hence, the CLE central charge as we defined it should correspond to the CFT central charge as it is defined in that context.

The introduction of the relative partition function in CLE and its relation to the one-point function of the stress-energy tensor are also interesting results of this work. In order to obtain a better understanding of this relative partition function, it is very instructive to conceptually connect it with partition functions of the underlying statistical model (or of CFT). It is also a goal of this subsection to clarify this connection. For simplicity of the discussion, domains $A,B,C,D$ will be simply connected domains (u6nless otherwise stated) excluding the point $\infty$.

\subsubsection{Interpretation of renormalised probabilities}

We start with an interpretation of the renormalised probability itself, $\Preg(\tou;A)_C$, definition \ref{preg}. Naturally, since the renormalised probability essentially requires that no loop intersects the boundary of $A$, one would expect that it is obtained from the number of configurations $Z_{C\setminus\cl{A}}^\tou$ in $C\setminus\cl{A}$ satisfying the conditions of $\tou$, and the numbers of configurations $Z_A$ in $A$ and $Z_C$ in $C$, through\footnote{I would like to thank D. Bernard for sharing this idea with me.} $Z_{C\setminus\cl{A}}^\tou Z_A / Z_C$. Of course, all these numbers are infinite in the scaling limit, and in fact so is this ratio. Hence, in order to have equality, we should normalise this ratio by another diverging number $N$. That is, we {\em multiplicatively renormalise} this ratio, where the regularised version is on the finite lattice, and the renormalisation is obtained by taking the scaling limit. In our definition of the renormalised probability, we took care in making the width of $\p A$ tend to zero in a precise way, depending on $A$. The renormalised probability may be made to equal the renormalised ratio of partition functions, but the diverging number $N$ in general will depend on $A$. Denoting it by $N_A$, we expect to have
\beq\label{PrenZ}
	\Preg(\tou;A)_C = N_A \frc{Z_{C\setminus\cl{A}}^\tou Z_A}{Z_C}.
\eeq
On the right-hand side, we implicitly understand that the scaling limit is taken. We expect $N_A$ to diverge in a non-universal way.

It is worth verifying that this expression agrees with some simple results that we found in the CLE context. Formula (\ref{ratiopreg}) can be derived straightforwardly from (\ref{PrenZ}):
\[
	\frc{\Preg(A)_B}{\Preg(A)_\hC} =
	\frc{Z_{B\setminus\cl{A}} Z_A}{Z_B} \frc{Z_\hC}{Z_{\hC\setminus\cl{A}} Z_A}
	= \frc{Z_{B\setminus\cl{A}} Z_{\hC\setminus \cl{B}} }{ Z_{\hC\setminus\cl{A}} } \frc{Z_\hC}{Z_B Z_{\hC\setminus\cl{B}}}
	= \frc{\Preg(\hC\setminus \cl{B})_{\hC\setminus \cl{A}}}{\Preg(\hC\setminus\cl{B})_\hC}.
\]
Also, the restriction property (\ref{eqrestreg}) follows immediately:
\[
	\frc{\Preg(\tou;A)_C}{\Preg(A)_C} =
	    \frc{Z_{C\setminus\cl{A}}^\tou Z_A}{Z_C} \frc{Z_C}{Z_{C\setminus\cl{A}} Z_A} =
		\frc{Z_{C\setminus\cl{A}}^\tou}{Z_{C\setminus\cl{A}}} = P(\tou)_{C\setminus \b{A}}.
\]

A less trivial result is the transformation property (\ref{fctf}). We may write
\beq\label{transfrat}
	\frc{\Preg(g \tou_C;g(A))_{g(C)}}{\Preg(\tou;A)_{C}} = \frc{N_{g(A)}}{N_A}
	\frc{Z_{g(C\setminus\cl{A})}^{g\tou} Z_{g(A)}}{Z_{g(C)}} \frc{Z_C}{Z_{C\setminus\cl{A}}^\tou Z_A}.
\eeq
In order to partly evaluate this, we need to know how the partition functions transform.

A conformal transformation of the domain of definition can be seen as a result of two steps: a reparametrisation of the initial domain, which obviously keeps the partition function invariant but changes the metric by an overall space-dependent factor, and a Weyl transformation that brings back the original metric, but under which the partition function transforms \cite{P81}. We use the standard setup where the trace of the bulk stress-energy tensor is zero, hence the metric we use is flat in the bulk (there is no trace anomaly, see for instance \cite{DFMS97}) -- it can be taken as the Euclidean metric. Then, we consider a partition function on $g(A)$ with that metric, and in the first step, we use $A$ as a parameter space for the domain $g(A)$. The metric it gives on $A$ (in the bulk) is obtained by $|dz|^2 \mapsto |dz|^2 |\p g(z)|^2$. In the second step, the Weyl transformation with a factor $e^{-\sigma(x)} = |\p g(z)|^{-2}$ brings the metric back to the Euclidean metric on $A$, and we have a partition function on $A$.

The transformation of the CFT partition function under a Weyl transformation was found by Polyakov in the context of random surfaces \cite{P81}: for $A$ any appropriate domain (say, any domain with piecewise smooth boundary), we have
\beq\label{transfoZ}
	Z_{g(A)} = e^{\frc{c}{48\pi} S_{\cl{A}}(\sigma)} Z_A
\eeq
where $c$ is the CFT central charge and $S_{\cl{A}}(\sigma)$ is the Liouville action of $\sigma$ on $\cl{A}$,
\beq\label{Liouville}
	S_{\cl{A}}(\sigma) =
	\int_{\cl{A}} d^2x\,\sqrt{\eta} \lt(\frc12 \eta^{ab} \p_a \sigma \p_b \sigma + R\sigma + \mu (e^{\sigma}-1)\rt).
\eeq
Here, $\eta^{ab}$ is the metric on $\cl{A}$ (and $\eta$ is its determinant), $R$ is the associated scalar curvature and $\mu$ is some UV-divergent, non-universal (i.e.\ lattice-model-dependent) scale. Our choice for $\eta^{ab}$ is the Kronecker delta $\delta_{ab}$ in the bulk of $A$.

In general, with curved boundaries, the curvature must have a non-zero contribution supported on the boundary. It is important that the integral in the Liouville action (\ref{Liouville}) covers the boundary of $A$ (which is the meaning of the notation $\int_{\cl{A}}$), so that it gets a non-zero contribution from this term. We will not need a precise description of the boundary term of the metric, but only some properties of the resulting contribution to the Liouville action. We will need that the contribution of the boundary $\p A$ to the Liouville action $S_{\cl{A}}(\sigma)$ only depends on the linear curvature along $\p A$ (besides the value of the function $\sigma$ on $\p A$). We will denote this contribution by $S_{\vec{\p} A}(\sigma)$, where $\vec{\p} A$ is the oriented boundary of $A$, counter-clockwise around the interior of $A$.

With this, we can now evaluate the ratio (\ref{transfrat}) (here, $A$ is again a simply connected domain):
\[
	\frc{N_{g(A)}}{N_A}
	\exp \frc{c}{48\pi}\lt[ S_{\cl{C}\setminus A}(\sigma) + S_{\cl{A}}(\sigma) - S_{\cl{C}}(\sigma) \rt]
	= \frc{N_{g(A)}}{N_A}
	\exp \frc{c}{48\pi}\lt[ S_{\vec\p (\hC\setminus A)}(\sigma) + S_{\vec\p A}(\sigma)\rt].
\]
On the right-hand side, only the parts of the Liouville actions supported on the boundary of $A$ remain. This expression makes it clear that the transformation property (\ref{fctf}) indeed involves a function $f(g,A)$ that may only depend on $g$ and $A$.

\subsubsection{Stress-energy tensor and the relative partition function}

We now turn to the relative partition function $Z(C|D)$, definition \ref{defiZ}. From (\ref{PrenZ}) it is expressed as
\beq
	Z(C|D) = \frc{Z_C Z_{\hC\setminus\cl{D}}}{Z_{C\setminus\cl{D}} Z_\hC}.
\eeq
Let us consider a transformation $g$ that is conformal on $\hC\setminus \cl{D}$, as well as the corresponding transformation $g^\sharp$ conformal on $C$ such that $g^\sharp(\p C) = g(\p C)$. As usual, we see $Z(C|D)$ as a function of $\p C$ and $\p D$, keeping $\p D$ on the component $C$ of $\hC\setminus \p C$. Let us consider the ratio
\beq
	\frc{Z(g^\sharp(C)|g(D))}{Z(C|D)} = \frc{Z_{g^\sharp(C)}}{Z_C} \frc{Z_{g(\hC\setminus\cl{D})}}{Z_{\hC\setminus\cl{D}}}
		\frc{Z_{C\setminus\cl{D}}}{Z_{g(C\setminus\cl{D})}}.
\eeq
We will first argue that this ratio is in fact independent of $\p D$, invariant under global conformal transformation, and, in some sense, universal. This is in agreement with both the global conformal invariance of renormalised probabilities (theorem \ref{invglob}), and with theorem \ref{theoex1pf}. Note that the latter theorem involves an infinitesimal conformal transformation on $\hC_w = \hC\setminus\cl{N(w)}$ for $w\in D$, which is indeed of the type of the transformation $g$ considered here. We will then provide further CFT arguments to show that the derivative $\Delta_w^{\hC_w}$ of this ratio reproduces the stress-energy tensor, in agreement with theorem \ref{theoex1pf}.

First, using the transformation property (\ref{transfoZ}), we find
\beqa
	\frc{Z(g^\sharp(C)|g(D))}{Z(C|D)} &=&
	\exp \frc{c}{48\pi} \lt[ S_{\cl{C}}(\sigma^\sharp) + S_{\hC\setminus D}(\sigma) - S_{\cl{C}\setminus D}(\sigma)\rt] \n
	&=& \exp \frc{c}{48\pi} \lt[ S_C(\sigma^\sharp) + S_{\hC\setminus \cl{C}}(\sigma)
		+ S_{\vec\p C}(\sigma^\sharp)-S_{\vec\p C}(\sigma)\rt]
	\label{alsp}
\eeqa
Note the careful inclusion/exclusion of domain boundaries in the Liouville actions. The last expression clearly is independent of $\p D$. Also, suppose $g$ is a global conformal transformation. Then we can choose $g^\sharp = g$ so that $\sigma^\sharp =\sigma$, and we are left with $\exp \frc{c}{48\pi} S_\hC(\sigma)$ (there is no boundary contribution). This is independent of $C$; that it should be 1 can then be obtained simply by sending $C\to\hC$ and $D \to \emptyset$. In order to argue that the right-hand side of (\ref{alsp}) is universal in some way, we need to argue that it is mostly independent of $\mu$ (the parameter in the Liouville action (\ref{Liouville})). Since $e^{\sigma} = |\p g|^2$, the $\mu$-terms in $S_C(\sigma^\sharp) + S_{\hC\setminus \cl{C}}(\sigma)$ can be combined into an integration over $\hC$ by change of coordinates; this then provides an overall factor that is independent of $\sigma$. This factor is seen to be 1 by setting $\sigma=0$ (that is, $g=\id$). As for the expression $S_{\p C}(\sigma^\sharp)-S_{\p C}(\sigma)$, there is a non-trivial metric on $\p C$, which we did not specify; but we expect that the resulting combination of $\mu$-terms is universal.

Second, we want to evaluate the derivative $\Delta_{w\,|\,\p C\cup \p D}^{\hC_w}$ of $\log Z(C|D)$ and show that it is the stress-energy tensor. Since this is the first derivative, the terms that are quadratic in $\sigma$ in the Liouville actions do not contribute. Also, as we argued above the bulk $\mu$-terms cancel out, and the bulk curvature terms are zero since the bulk metric is flat\footnote{There is a subtlety with the point at $\infty$ when the domain contains it: it takes all the curvature of the Riemann sphere. However, a careful calculation with the metric $d^2x/(1+|z|^2/R^2)^2$, where the curvature is re-distributed, shows that the limit $R\to\infty$ of the curvature term of the Liouville action gives zero contribution to the first derivative.}. This means that we are left only with the boundary contributions to the Liouville actions. Hence we find:
\beq\label{ZZ}
	\Delta_{w\,|\,\p C\cup \p D}^{\hC_w} \log Z(C|D) = \frc{c}{48\pi} \Delta_{w\,|\,\sigma}^{\hC_w}
		\lt[S_{\vec\p C}(\sigma^\sharp) - S_{\vec\p C}(\sigma)\rt]_{\sigma=0}.
\eeq
The term that is being differentiated is obviously invariant under small global conformal transformation, since we can then choose $\sigma^\sharp = \sigma$. This is important for the identification with the stress-energy tensor, since we need the transformation properties of the global holomorphic derivative (as discussed in subsections \ref{ssectdiff} and \ref{ssectCFT}).

Note that with an appropriate renormalisation of the partition function $Z_C^R$, we could guarantee that $S_{\hC\setminus \cl{C}}(\sigma) -S_{\vec\p C}(\sigma) = S_{\hC\setminus C}(\sigma)$ (that is, the boundary contributions simply get a minus sign for an opposite linear curvature of the boundary). Then, we would obtain
\beq\label{ZZ2}
	\Delta_{w\,|\,\p C\cup \p D}^{\hC_w} \log Z(C|D) = \frc{c}{48\pi} \Delta_{w\,|\,\sigma}^{\hC_w}
		\lt[S_{\cl{C}}(\sigma^\sharp) + S_{\hC\setminus C}(\sigma)\rt]_{\sigma=0}
		= \Delta_{w\,|\,\p C}^{\hC_w} \log (Z_C^R Z_{\hC\setminus \cl{C}}^R).
\eeq
On the right-hand side, we have not a single partition function, but a product. Again, this product guarantees that the derivative in directions of small global conformal transformations is zero. Yet, there is no ambiguity as to ``where'' the stress-energy tensor is inserted: the point $w$ must lie in $C$, and the analytic continuation of the function of $w$ that is obtained does not reproduce the derivative at points $w$ outside $C$. We will not need explicitly formula (\ref{ZZ2}).

Evaluating (\ref{ZZ}) directly would need a more precise understanding of the boundary terms in the Liouville actions. However, there is way of relating these boundary contributions to the stress-energy tensor without an explicit evaluation. Indeed, the stress-energy tensor may in fact be defined as the field generating the variation of the partition function under a change of metric $\eta \mapsto \eta+\delta \eta$ \cite{F84}:
\beq\label{basicform}
	\delta \log Z_A = \frc12 \int_{\cl{A}} d^2x\, \bra \delta\eta_{ab}(x) T^{ab}(x)\ket_A.
\eeq
Here, $A$ is some domain, and $T^{ab}$ is the symmetric stress-energy tensor in the canonical normalisation (in this normalisation, the charge $\int dx\, T^{0a}(x,y)$, in the quantisation on the line, generates $x^a$-derivatives with coefficient 1). With tracelessness $T^a_a=0$, it is related to the holomorphic and antiholomorphic components $T$ and $\b{T}$ via
\beq\label{Txx}
	T = -2\pi T_{zz} = -\pi (T_{xx} - i T_{xy}),\quad \b{T} = 2\pi T_{\b{z}\b{z}} = \pi (T_{xx} + i T_{xy}).
\eeq
This involves both a ``change of coordinates'' $z=x+iy,\,\b{z} = x-iy$, as well as a change of normalisation in order to guarantee the correct CFT normalisation of $T$ and $\b{T}$.

Under a transformation $g = \id + h$ that is conformal on the domain of definition, with $h$ small, the metric changes diagonally, $\delta\eta_{ab} = (\p h + \b\p \b{h}) \delta_{ab}$, so that we obtain the one-point function of the trace of the stress-energy tensor in (\ref{basicform}). This trace is zero except at the boundary, hence we are left with a boundary integration, as expected by the previous considerations. If we take $h(z) = \frc{\ep}{w-z}$ for some small complex $\ep$, we can evaluate $\Delta_w^{\hC_w} Z_{A}$ by extracting the part proportional to $\ep$ in $\delta Z_{A}$, and discarding the part proportional to $\b\ep$, as long as $w\not\in A$. If $w\in A$, we have to find a function $h^\sharp$ that has the same infinitesimal effect on $\p A$ but that is holomorphic on $A$. In this way, we could evaluate both terms on the right-hand side of (\ref{ZZ}): the first term by evaluating $\delta Z_C$ under $h^\sharp$, the second by evaluating $\delta Z_{C\setminus \cl{N(w)}}$ under $h$ and discarding the part that is integrated along $\p N(w)$.

Finding $h^\sharp$ in general is complicated. The simplest way to evaluate $\delta Z_C$ under $h^\sharp$ is rather to evaluate $\delta Z_{C\setminus \cl{N(w)}}$ under $h$ and take the limit where $N(w)\to \emptyset$ -- we just make a puncture at $w$. Evaluating the contribution of the puncture can be done via (\ref{basicform}), where the bulk metric change $\delta \eta_{ab}$ is singular at $w$, and not diagonal there. Denoting this contribution by $\delta Z_C[\mbox{puncture}]$, we simply find that
\[
	\frc{c}{48\pi}\lt.\Delta_{w\,|\,\sigma}^{\hC_w} S_{\vec\p C}(\sigma^\sharp)\rt|_{\sigma=0}
		= \frc{c}{48\pi}\lt.\Delta_{w\,|\,\sigma}^{\hC_w} S_{\vec\p C}(\sigma)\rt|_{\sigma=0} + \delta Z_C[\mbox{puncture}]
\]
and hence that
\beq\label{ZZ3}
	\Delta_{w\,|\,\p C\cup \p D}^{\hC_w} \log Z(C|D) = \delta Z_C[\mbox{puncture}].
\eeq
This formula quite directly leads to the one-point function of the stress-energy tensor (see below). In terms of the expression (\ref{ZZ2}), these considerations suggest that the product $Z_C^R Z_{\hC\setminus \cl{C}}^R$ takes care of the boundary conditions, upon inserting the bulk stress-energy tensor, by a ``method of images.'' Also, we see that the presence of the domain $D$ in the CLE relative partition function $Z(C|D)$ has two important effects: it cancels the boundary contributions to the singular metric change, so that only the puncture contribution remains, and it cancels out the non-universal numbers $N_C$ involved in relating renormalised probabilities to partition functions, (\ref{PrenZ}).

The calculation of $\delta Z_C[\mbox{puncture}]$ goes as follows. In general, for a transformation of coordinates $\delta x^a = v^a(x,y)$, the metric change is $\delta \eta_{ab} = \p_a v_b + \p_b v_a$. In our case, we simply have $\delta z = h(z)$, so that
\[
	\p_x v_x + \p_y v_y = \p h + \b\p \b{h},\quad \p_x v_x - \p_y v_y = \b\p h + \p \b{h},\quad
		\p_x v_y + \p_y v_x = -i(\b\p h - \p \b{h}).
\]
Using the formulas \cite{F84}
\[
	\frc{\p}{\p z} \frc{1}{w-z} = \frc{\p}{\p \b{z}} \frc{1}{\b{w}-\b{z}} = -\pi \delta^2(z-w)
\]
it is straightforward to arrive at
\[
	\delta \eta_{ab} T^{ab} = -2\pi \delta^2(z-w) \lt( (\ep+\b\ep) T_{xx} - i (\ep-\b\ep) T_{xy} \rt).
\]
Hence, using (\ref{Txx}) and (\ref{basicform}) and keeping the $\ep$ part only we indeed obtain
\[
	\Delta_{w\,|\,\p C\cup \p D}^{\hC_w} \log Z(C|D) = \bra T(w)\ket_C.
\]

Note that we could as well have used formula (\ref{basicform}) combined with the calculations found in the proofs of theorems \ref{theowardplane} and \ref{theowarddom} in order to reproduce the stress-energy tensor from CFT arguments. However, the derivation above, using a singular metric, is simpler and provides an alternative route in the CFT context.

Note finally that the expression $Z_C^R Z_{\hC\setminus \cl{C}}^R$ found in (\ref{ZZ2}) seems to be closely related to the renormalised probability $\Preg(C)_\hC$ itself, or to $\Preg(\hC\setminus \cl{C})_\hC$, according to (\ref{PrenZ}). We could perhaps have obtained the latter directly from the relative partition function $Z(C|D)$ under the derivative $\Delta_w^{\hC_w}$, by using the fact that the derivative is independent of $D$. Indeed, it could be argued that the derivative of the denominator in definition \ref{defiZ} tends to something that is independent of $C$ as $D\to C$, so could simply be omitted. However, we have not proven this, and there may be subtleties having to do with the renormalisation, in particular with the non-universal factor $N_A$ involved in (\ref{PrenZ}).

\subsection{Loops and particle world-lines}

Recall that there are (at least) three natural physical interpretations of the stress-energy tensor, depending on the point of view that we take about QFT: it describes metric changes (in two-dimensional statistical models), it groups currents associated to space and time translation symmetries (in quantum chains), or it measures the flow of energy and momentum of relativistic particles (in models of relativistic quantum particles travelling in one dimension). The previous subsection makes it clear that the global holomorphic derivative $\Delta_w^{\hC_w}$ corresponds to a singular metric change, hence this indicates agreement with the interpretation of the stress-energy tensor as a field describing metric changes. On the other hand, the conformal Ward identities themselves, as explained in subsection \ref{ssectCFT}, point to the interpretation as currents associated to symmetries. In order to understand the third interpretation in the context of CLE, we would need to understand the relation with relativistic particle world-lines (trajectories in space-time). We do not fully understand this yet, but we may provide some heuristic ideas about how it could work. In fact, this gives us a way of understanding the particular form of the stress-energy tensor that we obtained here (and in \cite{DRC}): that of a renormalised probability for a spin-2 rotating ellipse.

The interpretation of the components $T_{xx}$, $T_{xy}$, $T_{yx}$, $T_{yy}$ of the stress-energy tensor are straightforward in terms of a ``gas'' of particles, with $y$ the (imaginary) time and $x$ the one-dimensional space: the component $T_{yy}$ measures the energy density; the off-diagonal components, equal to each other, measure the energy current or the momentum density; and the component $T_{xx}$ measures the momentum current. In this interpretation, we naturally have $T_{xy} = T_{yx}$, but the tracelessness relation $T_{xx}+T_{yy}=0$ is a strong statement about the dynamics of the particles. For clarity, let us keep $T_{xx}$ and $T_{yy}$ unrelated. The expression (\ref{Txx}) of the holomorphic stress-energy tensor in terms of Euclidean components then becomes
\beq\label{Txx2}
	T = \frc{\pi}2 (-T_{xx} + T_{yy} + i T_{xy} + i T_{yx}).
\eeq

\begin{figure}
\bc
\includegraphics[width=5cm,height=5cm]{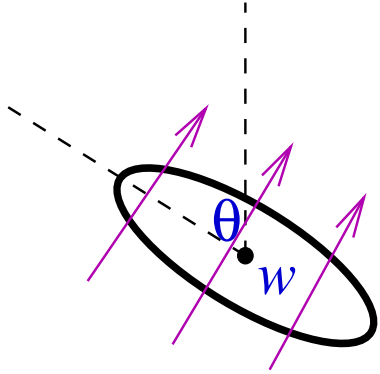}
\ec
\caption{The elliptical domain at angle $\theta$, and parts of world-lines crossing it.}
\label{figlines}
\end{figure}

The main idea behind the world-line interpretation is that we should construct world-lines that are {\em perpendicular} to the CLE loops, and give them a direction towards increasing $y$ (so that particles travel forward in time). Then, the renormalised probability $\Preg(\tou;E(w,\ep,\theta))_C$ should be a measure of the density of world-lines in the direction perpendicular to the principal axis of the ellipse (since the CLE loops around and inside the elliptical domain should tend to align with the ellipse) -- see figure \ref{figlines}.
More precisely, it should measure the density of the {\em components} of the world-lines in that direction. At $\theta=\pi/2$, the world-line direction is vertical, so we are simply measuring the energy density, $T_{yy}$. For this value of $\theta$, definition \ref{T} gives a coefficient 1. At $\theta = \pi/4$ and $\theta=3\pi/4$, the world-line directions are diagonal, going towards the right and the left, respectively. Hence, we are measuring momentum densities in different directions, $T_{xy}$ and $-T_{xy}$ respectively, and for these values of $\theta$, definition \ref{T} gives coefficients $i$ and $-i$ respectively. Finally, at $\theta=0$, the world-line direction is horizontal. This is not as evident, because it formally represents particles going ``faster than the speed of light'' (although we are in Euclidean signature). In order to assess this case, we could simply take a different time direction, towards increasing $x$ instead of $y$ (with the space direction towards decreasing $y$). Then, we are measuring $T_{xx}$, and definition \ref{T} gives a coefficient $-1$. Note that the diagonal cases could also have been done with this different time direction, giving the same results. Hence, analysing four fixed values of $\theta$ in the integrand in definition \ref{T}, and interpreting the probability through particle world-lines, we reproduced the four terms on the right-hand side of (\ref{Txx2}) (up to an overall positive normalisation).

Interestingly, these heuristic ideas suggest that definition \ref{T} proportional to the component $T_{zz}$ of the stress-energy tensor could hold as well in loop models that do not possess conformal invariance (that is, more general QFT models, expressed in terms of fluctuating loops).

Finally, it is worth mentioning that this point of view shares many properties with a recent construction of P. Mansfield of the electric and magnetic fields based on Faraday's lines of force and on a generalisation of this to surfaces \cite{Mans}\footnote{I am grateful to P. Mansfield for sharing with our group at Durham his ideas about Faraday's lines of force, and for letting me see the manuscript before publication.}. That paper constructs, in particular, the solution to Maxwell's equations in the case of two static opposite-charge particles, by assuming that there is a line between them where all the electric field is concentrated (and tangential), and by averaging over fluctuations of this line according to a certain measure. In two dimensions, these Maxwell's equations are equivalent, for the component $E := (E_x-iE_y)/2$ of the electric field vector ${\bf E}$, to holomorphy $\b\p E=0$ except for two points with prescribed singularity structures (simple poles of opposite residues where the charges are). Hence, Mansfield's construction (adapted to two dimensions) is a probabilistic solution to this analytical problem. It is similar to our construction of the stress-energy tensor here, and perhaps more clearly in the SLE case in \cite{DRC}. Indeed, we consider a vectorial (i.e.\ complex) random variable supported on random curves with a direction (i.e.\ phase) determined by the local direction of the curves, and whose average solves a condition of holomorphy except for points with prescribed singularity structures. Of course, our case is slightly different in that the resulting object has spin 2. It would be worth exploring further the relation between these constructions.

\subsection{Objects, CFT fields, and duality}\label{ssectobjects}

The definition \ref{object} of objects is likely to be general enough to include the scaling limit of all useful local statistical variables. Then, the correlation functions as defined in definition \ref{object} are expected to reproduce the scaling limit of their averages, in accordance with the usual wisdom of QFT. For instance, the scaling limit of the spin variable at $w$ in the Ising model should correspond to the limit $\ep\to0$ of the random CLE variable that is positive if the number of loops separating a disk of radius $\ep$, centered at $w$, from the boundary of the domain is even, and negative otherwise. This can clearly be written as an object in the set-up of definition \ref{object}. It should be noted that this is not supported at $w$; rather, as for the variable $n(z_1,z_2)$ above, the support is $\hC$, since the boundary of the domain is used in order to know the parity. The product of two spin variables can be identified with a similar random variable, but where we look for the parity of the number of loops separating the two spins, at points $w_1$ and $w_2$ for instance (with small disks around them). This is supported on $\{w_1,w_2\}$. Another field that can be formed out of these is the energy field of the Ising model. Since it occurs in the operator product expansion (OPE) of two spin fields, according to the standard arguments of CFT (see, for instance, \cite{Gins,DFMS97}), it can be written as the limit $w_1\to w_2$ of some integral over $w_1$ and $w_2$ around a point $w$ of the product of spin fields at $w_1$ and $w_2$. This is then supported on the point $w$.

The examples of objects above show that some are quite different from the stress-energy tensor, since they are not supported on a point. As is clear from the sketch of the proof of the universality statement, any object supported on a point is determined by the local behaviour of loops around this point only; we will say that it is {\em local with respect to the loops}. In particular, such objects can be evaluated in CLE configurations by only looking at a loop surrounding this point, and since there is always one that separates it from other objects (supported away from that point) in the correlation functions, it is sufficient to know the one-point averages in all possible domains. This is not true of objects supported on an extended set.

There are few statistical variables that can be expected to have, in the scaling limit, this ``single-point support'' property. We would like to propose that the only group of fields with this property is that of all fields generated, through the operator algebra, by the identity sector and by the energy-field sector. The identity sector is composed of (besides the trivial identity field) the stress-energy tensor and all its descendants. We have seen that the stress-energy tensor is indeed supported on a point. Its descendants under the Virasoro algebra are obtained by considering short-distance expansions with the stress-energy tensor itself, so are also supported on a point. The physical interpretation is that the sector of the stress-energy tensor measures local properties of particle trajectories, as seen above, hence should be local with respect to these trajectories. On the other hand, the energy field is the field corresponding to the hamiltonian density of the statistical model, which determines the Boltzmann weight of the configurations. There is always such a field, and, in a local statistical model, it reads only local variables of the statistical model. Since the loops should represent the lines where a ``defect'' is present, only on these lines the energy field should read a non-zero value. This is of course particularly clear in the Ising model. Hence, the energy field, as we saw in the Ising case, should also be local with respect to the loops, and likewise, its descendants under the Virasoro algebra are also local. 

These considerations as well as those of the previous subsection point to the dual nature of the random loop construction of CFT. Indeed, loops are both characteristic of the configurations of an underlying statistical model, and of the particle trajectories in a Feynman-type construction of CFT. Associated to these two interpretations, there are two families of fields that are local with respect to the loops. They measure the corresponding two types of fluctuating energies: the statistical energy, and the particle energy.

\subsection{Perspectives}

Perhaps the most pressing calculation is that relating the central charge as we defined it, to the parameter $\kappa$ characterising the CLE measure, or to the time of the Poisson process involved in the stochastic construction of CLE (whose relation to $\kappa$ is known) \cite{W05a,ShW07}. We have given strong arguments that our central charge is the correct one, through CFT considerations, but it would be very interesting to provide a CLE proof that indeed we find the expected formula,
\[
	c = \frc{(6-\kappa)(3\kappa-8)}{2\kappa}.
\]

Many extensions of this work are possible. First, the antiholomorphic component of the stress-energy tensor $\b{T}$ can of course be constructed without difficulties along entirely similar lines. But also, it would be very interesting to develop the whole identity sector through Fourier components of renormalised probabilities of similar geometric figures. This should be possible, because all fields in the identity sector are local with respect to the loops, and are, in a sense, of ``geometric character.'' Second, it is possible quite straightforwardly to extend the applicability of the conformal Ward identity to other objects than simple CLE events. For instance, for the Ising spin, as we discussed, the object should simply be, loosely speaking, the limit $\ep\to0$ of an appropriately normalised random variable evaluating the parity of the number of loops outside a small disk of radius $\ep$ (see subsection \ref{ssectobjects}) (one can imagine many other objects characterised by a small disk in a similar way). The normalisation should simply make the object a primary field of a given dimension and zero spin. Other constructions, taking Fourier transforms for instance, will lead to non-zero spins, and eventually to non-primary transformation properties. Knowing the transformation properties of an object, the derivations of theorems \ref{theowardplane} and \ref{theowarddom} can be repeated, and immediately lead to the correct conformal Ward identities. In particular, since we have proven the transformation properties of the stress-energy tensor itself, this gives the conformal Ward identities with multiple insertions of the stress-energy tensor. The chief complication involved in such general conformal Ward identities is that of the existence of the limits involved, and of the independence on the order of the limits.

From multiple stress-energy tensor insertions, we may develop the basis for the algebraic setup of CFT. Indeed, standard arguments of CFT gives rise to the Virasoro algebra, with the central charge equal to the one that occurs in the transformation property of the stress-energy tensor. From this, the Virasoro vertex operator algebra is obtained essentially once we have proven the correct analytic properties of the insertions of fields in the identity sector. We hope to develop this in a future work.

The construction of the {\em boundary} stress-energy tensor and its descendants is also possible by similar methods, as was done in the SLE context in \cite{DRC}. In fact, this does not require many of the assumptions that we had to make in the present work, since it does not require probability functions on $\hC$ and on doubly connected domains. Also, in this context, we may justify our choice of the events $\ev(A,\varep,u)$ for defining renormalised probabilities. The main property of these events is that loops in $A$ and outside $A$ are ``separated.'' Such a separation can also be done by other events, for instance, asking that at least one loop be present that surrounds $A$ in the corridor along $\p A$ of thickness defined by $\varep$ and $u$. Many (but not all) of the calculations and proofs in the present work (this paper and the previous one \cite{I}) could have been done with such events instead. However, when considering the boundary case, we need a similar event, but, instead of the closed loop $\p A$, we must consider an arc starting and ending on the boundary of the domain of definition. In this case, there is no possibility other than the event asking that no loop cut through the fattened arc.

Certainly, it would be important to prove that the assumptions that were used in this and our previous paper \cite{I} hold in CLE for $8/3<\kappa\leq 4$; notably the symmetry property that allows us to construct probability functions on $\hC$ and on doubly connected domains, and differentiability.

Finally, extensions beyond CFT with central charges in the range $0<c\leq 1$ and to higher dimensions would be very interesting, although they necessitate a proper understanding of fluctuating objects. For $c>1$ CFT, it should be possible to construct other symmetry fields in a similar fashion (likewise, in fact, for the free boson $c=1$ CFT, where a $U(1)$ symmetry field exists). For other situations, the stress-energy tensor is likely to be a good starting point for studying local fields, as it possesses many simplifying properties.

\end{document}